\documentclass[10pt,conference]{IEEEtran}\IEEEoverridecommandlockouts
\usepackage{amsmath,amssymb,amsfonts}
\usepackage{hyperref}
\usepackage{algorithmic}
\usepackage{graphicx}
\usepackage{textcomp}
\usepackage[normalem]{ulem}
\usepackage{array}

\def\BibTeX{{\rm B\kern-.05em{\sc i\kern-.025em b}\kern-.08em
    T\kern-.1667em\lower.7ex\hbox{E}\kern-.125emX}}
    
\usepackage[dvipsnames]{xcolor}

\usepackage{tabularx}
\usepackage{booktabs}
\usepackage{multicol}
\usepackage{multirow}
\usepackage{soul}

\usepackage{stfloats}

\usepackage{etoolbox}
\makeatletter
\patchcmd{\@makecaption}
  {\scshape}
  {}
  {}
  {}
\makeatother

\graphicspath{{Figures/}}

\usepackage[
backend=biber,
style=ieee,
sorting=none
]{biblatex}

\hypersetup{
    colorlinks=true,
    linkcolor=Blue,
    filecolor=magenta,      
    urlcolor=Blue,
    citecolor=Blue,
    pdftitle={Context Conquers Parameters: Outperforming Proprietary LLM in Commit Message Generation}
}
\usepackage{wrapfig}

\usepackage[many]{tcolorbox}    	% for COLORED BOXES (tikz and xcolor included)
\definecolor{main}{HTML}{31363F}    % setting main color to be used
\definecolor{sub}{HTML}{EEEEEE}     % setting sub color to be used
\tcbset{
    sharp corners,
    colback = white,
}  
\newtcolorbox{boxH}{
    colback = sub, 
    colframe = main, 
    boxrule = 0pt, 
    leftrule = 5pt % left rule weight
}
\newtcolorbox{boxA}{
    fontupper = \bf,
    boxrule = 1pt,
    colframe = main % frame color
}
\usepackage{subfiles} % Best loaded last in the preamble

\begin{document}

\title{Context Conquers Parameters: Outperforming Proprietary LLM in Commit Message Generation}

% \title{}

% \title{ECOMEG: Efficient Commit Message Generator}

\author{\IEEEauthorblockN{Aaron Imani}
\IEEEauthorblockA{
\textit{University of California, Irvine}\\
Irvine, USA \\
aaron.imani@uci.edu}
\and
\IEEEauthorblockN{Iftekhar Ahmed}
\IEEEauthorblockA{
\textit{University of California, Irvine}\\
Irvine, USA \\
iftekha@uci.edu}
\and
\IEEEauthorblockN{Mohammad Moshirpour}
\IEEEauthorblockA{
\textit{University of California, Irvine}\\
Irvine, USA \\
mmoshirp@uci.edu}
}

\maketitle

% For page number
% \thispagestyle{plain}
% \pagestyle{plain}

\begin{abstract}
Commit messages provide descriptions of the modifications made in a commit using natural language, making them crucial for software maintenance and evolution. Recent developments in Large Language Models (LLMs) have led to their use in generating high-quality commit messages, such as the Omniscient Message Generator (OMG). This method employs GPT-4 to produce state-of-the-art commit messages. However, the use of proprietary LLMs like GPT-4 in coding tasks raises privacy and sustainability concerns, which may hinder their industrial adoption. Considering that open-source LLMs have achieved competitive performance in developer tasks such as compiler validation, this study investigates whether they can be used to generate commit messages that are comparable with OMG. Our experiments show that an open-source LLM can generate commit messages that are comparable to those produced by OMG. In addition, through a series of contextual refinements, we propose lOcal MessagE GenerAtor (OMEGA)
, a CMG approach that uses a 4-bit quantized 8B open-source LLM. OMEGA produces state-of-the-art commit messages, surpassing the performance of GPT-4 in practitioners' preference.

\end{abstract}


\section{Introduction}

% Commit message is important for software maintenance and researchers have been trying to use recent LLM's adavancement to automatically generate it.
Commit Messages (\textbf{CM}) play a crucial role in documenting changes in version control systems, facilitating maintenance and evolution of the software 
% \cite{Barnett2016TheGitHub, Hindle2009AutomaticCategories, Rebai2020RecommendingAnalysis}. 
\cite{Rebai2020RecommendingAnalysis}. 
Researchers have leveraged Natural Language Processing techniques to develop automatic methods for Commit Message Generation (\textbf{CMG}), with the goal of improving the quality of human-written messages \cite{Tian2022WhatMessage, li2024omg}. In addition to detailing the changes (``What'' information) and their rationale (``Why'' information) \cite{Tian2022WhatMessage}, researchers have identified additional expected criteria from practitioners' perspective \cite{li2024omg}.
Thus, CMG can be regarded as a complex reasoning task requiring a comprehensive and precise understanding of the commit context, as well as the ability to formulate a CM that meets all the quality criteria developers expect.

Given this complexity, CMG methods have significantly advanced with the introduction of Large Language Models (\textbf{LLM}) trained using reinforcement learning from human feedback (RLHF) \cite{ziegler2020finetuninglanguagemodelshuman}, such as GPT-4 \cite{gpt4}. The enhanced reasoning capabilities of these models have resulted in state-of-the-art performance in CMG \cite{li2024omg, Eliseeva2023}.
% Notably, the introduction of Large Language Models (\textbf{LLM}) trained with reinforcement learning from human feedback (RLHF) \cite{ziegler2020finetuninglanguagemodelshuman}, such as GPT-4 \cite{gpt4}, has encouraged researchers in the field of software engineering to utilize them in order to produce state-of-the-art for various tasks, including in Commit Message Generation (\textbf{CMG}) \cite{zhang2024automatic-cmg-crititical-review, wu2024-cmg-chatgpt}.
%\hl{\textbf{Done.} the following sentence in this para is not connected to the first part of }
Notably, Li et al. utilized GPT-4, a proprietary LLM with 1.76 trillion parameters \cite{GPT-4Wikipedia}, to outperform traditional state-of-the-art CMG approaches \cite{li2024omg}. They introduced Omniscient Message Generator (\textbf{OMG}), a ReAct \cite{Yao2022ReAct:Models} agent that produces high-quality CMs by leveraging six contextual pieces of information about a commit. GPT-4 is employed in OMG to generate three out of the six contextual pieces of information, and it also serves as the reasoning engine in the ReAct chain.
Although OMG produces CM with state-of-the-art quality, its reliance on a proprietary LLM introduces certain organizational and environmental risks.
% They found that using the commit diff alone is insufficient to meet practitioners' expectations for high-quality commit messages. OMG leverages six different contextual pieces of information about a commit: 1) Associated issues on the version control system or issue tracking service 2) Associated pull requests 3) Relative importance of changed files 4) The software maintenance activity type of the commit 5) Changed methods' summaries 6) Changed classes' summaries. 
% In addition to the ReAct loop and the generated CM, the latter three components of the commit context are generated by a large language model (\textbf{LLM}). 
% OMG employs OpenAI's GPT-4 \cite{gpt4} to generate these LLM-derived elements of the commit context as well as the reasoning engine in its ReAct chain.
% GPT-4 is utilized in the OMG to generate three out of six contextual information about a commit as well as the reasoning engine in its ReAct chain.

% Proprietary LLMs propose risks and OMG uses them. So, it also introduces the same risks, which limits its applicability in the industry.
%\hl{\textbf{Done.} This para is not connected with the previous para. Why are we suddenly talking about the problems of commercial LLM}

Third-party proprietary LLM APIs have been introducing privacy risks for companies
% \cite{Shashidhar2023DemocratizingModels, SamsungLeak, AmazonsWarn} 
\cite{SamsungLeak, AmazonsWarn} 
and research has been done to investigate LLMs' privacy and security implications 
% \cite{Wu2024ASystems, Yao2024AUgly}. 
\cite{Yao2024AUgly}. 
For instance, a developer at Samsung accidentally shared sensitive internal source code with ChatGPT \cite{openai2024chatgpt}, leading Samsung to ban the use of all proprietary chatbots due to the difficulty in accessing and deleting shared information \cite{SamsungLeak}.
OMG requires sharing sensitive source code information, including the bodies of affected methods and classes before and after the commit, with a third-party API (GPT-4). This introduces privacy risks, making its adoption by the industry problematic and limiting its practical application.

% \hl{\textbf{How about this revised para?}this sentence is nit doing it..Is size correlated with energy consumption?}
% The adoption of proprietary LLMs leads to carbon emissions that are several times greater than those produced by training such models each year\cite{Chien2023Reducing2035}. 
Additionally, researchers have shown that making requests to proprietary LLMs can lead to annual carbon emissions greater than the emission during training such LLMs \cite{Chien2023Reducing2035}. 
OMG makes 12 LLM requests (4 in preparing commit context + 7 Thoughts in ReAct (number of available tools to the Agent) + 1 Initial ReAct Thought)
% \footnote{12 = 4 in preparing commit context + 7 Thoughts in ReAct (number of available tools to the Agent) + 1 Initial ReAct instruction)} 
to generate a CM.
A study on 24 popular Java repositories has shown that on average, a developer pushes 3 commits daily \cite{Ferreira2023}.
Assuming this number holds in industrial cases, a company with 1,000 developers using OMG would generate over a million CMs for a single project annually, resulting in 12 million calls to a proprietary LLM.
% OMG sends multiple requests to a proprietary LLM in preparing parts of commit context. In addition, the prompting approach adopted by OMG, ReAct \cite{Yao2022ReAct:Models}, requires sending multiple requests to produce a CM. 
This makes the wide adoption of OMG a sustainability threat that introduces environmental risks.
% \hl{LOVE THIS}
% The token-intensive prompting approach (ReAct) \cite{xia2024agentlessdemystifyingllmbasedsoftware} used by OMG, along with its reliance on a proprietary LLM to generate parts of the commit context, renders it an unsustainable CMG approach in the long run. 
These limitations in OMG motivate the need for a shift away from a proprietary LLM-based approach.

% Analyzing 24 popular Java-based projects on GitHub, revealed an average of 3 commits per day per developer. Hence, a small development team with 50 developers would make 4500 commits per month. Given the necessity of GPT-4 in summarizing changed classes, changed methods, and producing the CM through a token-intensive prompting approach (ReAct) \cite{xia2024agentlessdemystifyingllmbasedsoftware}, such a team would generate 1.8

% Open-source LLMs don't have the same problems as the proprietary ones and SE researchers have been trying to use it in solving SE tasks. Since no one has tried it for CMG, we want to explore it.
Open-source LLMs (\textbf{OLLM}) offer a cost-effective and privacy-conscious alternative, as they mitigate the privacy concerns associated with proprietary models. Deployable on local GPUs due to their smaller number of trained parameters, OLLMs are more sustainable for adoption in LLM-based automations \cite{Faiz2023LLMCarbon:Models}.
% A para about the success and failures of ollms in se tasks
Software engineering researchers have attempted to use OLLMs to address several development tasks. While they have shown promising results in some areas 
% \cite{Valero-Lara2023ComparingGeneration, Jimenez2023SWE-bench:Issues}, 
\cite{liu2024loganalysis, zhong2024chatgptreplacestackoverflowstudy},
a performance gap has been noted when compared to proprietary LLMs 
% \cite{Chen2023ChatGPTsUp}. 
\cite{yin2024multitaskbasedevaluationopensourcellm, pan2024codetranslation}.

However, with the advent of new OLLMs that achieve performance on par with proprietary LLMs across various benchmarks and leaderboards 
% \cite{Liu2023IsGeneration, Liu2024RepoQA:Understanding}, 
\cite{Liu2024RepoQA:Understanding}, 
the likelihood of their successful adoption for a wide range of tasks is increasing. 
% \cite{codeqwen1.5, llama3}. 
%\hl{\textbf{Done.} why is this work important is not coming across. You need to merge the para with the prior para and have a clear message which is: OLLM is important due to various issues of commercial LLM. Not a lot of SE people has adopted OLLMs, not in CMG for sure. We take the first step and it is especially important due to the context-dependent nature of CMG, as Jiawei et al.'s approach requires multiple queries and such. And every query is equivalent of having one 15 watt light bulbe on for a year and also monetary issue making it difficlt/ So we do this study.}
% \hl{\textbf{Restructured the para.} bad start of para}
% None of the 
Since prior work has not investigated the application of OLLMs in generating CMs that meet practitioners' expectations and are able to achieve performance similar to the state-of-the-art CMG technique \cite{li2024omg}, 
% Therefore, 
our study aims to take the first step in replacing GPT-4 in OMG with an OLLM and to assess the feasibility of generating CMs comparable to those produced by OMG. To address this goal, we formulated our first research question.

% rq1
\textbf{RQ1: Can an OLLM generate CMs comparable to a state-of-the-art LLM (GPT-4)?}

% We used an OLLM and could generate comparable CMs to OMG. However, it lacked comprehensiveness.
To address our first research question, we experimented with using an OLLM instead of GPT-4 to generate LLM-derived commit context and produce high-quality CMs. 
We utilized automated machine translation evaluation metrics and practitioner surveys to measure the quality of the generated CMs.
%\hl{\textbf{Done.} in the previous sentence, you're giving away the result. Not usually a good idea but might be ok in this case}
%Section III details our experiments and survey methodology, while Section IV presents the survey results.

Based on our results, although the OLLM produced CMs with quality comparable to OMG in various aspects, it did not meet practitioners' expectations in one key aspect: Comprehensiveness \cite{li2024omg}. This indicates that the OLLM-generated CM missed details that were covered by OMG. To address this shortcoming, we propose our second research question.

%Based on our results, although the OLLM was able to produce CMs with quality comparable to OMG, it did not meet practitioners' expectations in one quality aspect: Comprehensiveness. This means that the OLLM-generated CM missed details that were covered by OMG.
% we found that the CMs generated by the OLLM were less comprehensive than those produced by OMG, missing details that OMG covered. 
%\hl{\textbf{Done.} the tone of the prior sentence should be, although we were able to produce good quality CM, they lacked one quality aspect....}
%Hence, we aimed to bridge this identified performance gap by proposing our second research question.
%\hl{\textbf{Done.} You have defined OLLM early on so you should not write open source LLM anymore anywhere}

\textbf{RQ2: How can we bridge the comprehensiveness gap between the CMs produced by an OLLM and the CMs generated by a state-of-the-art LLM (GPT-4)?}

% We introduced CMMS and an additional preprocessing step to answer RQ2.
To answer this RQ, we introduced \textbf{C}hange-Based \textbf{M}ulti-Intent \textbf{M}ethod \textbf{S}ummarization (\textbf{CMMS}) to provide refined contextual information that bridge the comprehensiveness gap by refining the commit context. Through automated evaluation and a second survey, we examined the refined context's effectiveness in addressing RQ2.

While the improved prompt with the enhanced commit context leads to CMs even closer to those generated by OMG, the running the OLLM for inference requires at least an NVIDIA A6000 GPU with 48GB of VRAM. This high resource demand can limit the practical usefulness of our CMG approach for individual developers or smaller teams with limited budgets. 
Furthermore, adopting an OLLM with fewer trained parameters results in a more sustainable CMG approach by minimizing its carbon emission \cite{Faiz2023LLMCarbon:Models}, aligning with one of the primary objectives of this study.
Thus, we aimed to explore the possibility of generating comparable CMs using a smaller OLLM (\textbf{SLM}) that can run on a local GPU with as little as 8GB of VRAM \cite{LlamaParseur}, \textbf{reducing the GPU VRAM requirement by 84\%}. This investigation is formulated as the third research question of this study.

\textbf{RQ3: Can a smaller OLLM (SLM) produce CMs comparable to a state-of-the-art LLM (GPT-4)?}

% We tried the same prompt and context with SLM but it didn't work well. Then we proposed two diff augmentation techniques to fix it and answer RQ3.
%\hl{\textbf{Done.} same issue with the following para, too much details, rewrite}
To address this research question, we initially employed the same prompt and context as we used to answer RQ2 in an attempt to achieve comparable results. However, as anticipated, the automated scores for the generated CMs by our tested SLMs were considerably lower compared to those produced by the larger OLLM.
The poor scores, despite the enhanced context, led us to hypothesize that the SLM is not capable of correctly understanding the underlying changes in a diff. Having validated our hypothesis through an analysis study, we developed two commit diff augmentation techniques, Diff Narrator and \textbf{F}ine-grained \textbf{I}nteractive \textbf{D}iff \textbf{EX}plainer \textbf{(FIDEX)}, designed to clarify the changes in a diff. As with previous research questions, we used automated metrics and a third practitioner survey to evaluate the effectiveness of our diff augmentation techniques in closing the gap between the commit messages generated by the SLM and those produced by the OLLM.

In summary, our study makes the following contributions:

\begin{enumerate}
%\item We explore the feasibility of generating state-of-the-art CMs that meet practitioners' quality criteria using an OLLM.
\item We demonstrate that replacing GPT-4 with an OLLM using the same commit context as OMG produces comparable CMs in all human CM quality evaluation criteria except comprehensiveness. 

\item We introduce a new method summarization approach called \textbf{C}hange-based \textbf{M}ulti-Intent \textbf{M}ethod \textbf{S}ummarization (\textbf{CMMS}) for software engineering tasks that rely on code changes. %Using this technique along with an additional preprocessing step, we bridged the identified comprehensiveness gap.

    % \item We demonstrate that high-quality CMG is feasible using an OLLM with zero-shot prompting. The effectiveness of the generated CMs is validated through practitioner surveys and automated machine translation metrics.
    % \item We address the performance gap in CMs generated by the OLLM by adding a preprocessing step when summarizing affected classes and methods and by proposing Change-based Multi-Intent Method Summarization (CMMS). 
    % \hl{\textbf{Revised the bullet point. }so adding a preprocessing step is our contribution!}
    \item We propose two augmentation techniques for commit diff, Diff Narrator and \textbf{F}ine-grained \textbf{I}nteractive \textbf{D}iff \textbf{EX}plainer \textbf{(FIDEX)} that boost SLM's performance in the CMG.
    \item We propose the state-of-the-art CMG approach, l\textbf{O}cal \textbf{M}essag\textbf{E} \textbf{G}ener\textbf{A}tor (\textbf{OMEGA}), that employs a 4-bit quantized SLM with 8B trained parameters that runs on a local GPU with as little as 8GB VRAM to generate CMs that are preferred by practitioners over those generated by OMG.
    % \hl{\textbf{Rephrased to better reflect our positive results.} CM that outperform OMG in all human CM quality criteria except conciseness. TRUE?}
    
    % \item We show that our contextual refinements enable a 4-bit quantized SLM with 8B trained parameters that runs on a GPU with as little as 8GB VRAM to generate CM that outperform OMG in all human CM quality criteria.
    
    % that is an LLM-based approach to highlight the differences between the old and new vesrions of affected Java files in a commit. Augmenting the raw diff with the produced diff summary by IDS enables a 4-bit quantized SLM with 8B trained parameters to produce commit messages comparable to those generated by OMG using GPT-4.
    
    % \item For the first time, we explore the applicability of MAD prompting in the context of commit message generation. By examining different settings, we introduce a MAD-based approach that enables a quantized OLLM with 8 billion parameters to produce CMs comparable to those generated by OMG using GPT-4.
\end{enumerate}

The remainder of this paper is structured as follows. In \hyperref[related-work]{Section II}, we review related research pertinent to our work. In \hyperref[method]{Section III}, we detail our methodology for addressing all research questions. \hyperref[results]{Section IV} presents the results of our experiments and surveys. 
% \hyperref[implications]{Section V} describes the implications of our work for the researchers and for the industry.
In \hyperref[threats]{Section V}, we highlight potential threats to the validity of our findings and the measures taken to mitigate them. Finally, \hyperref[conclusion]{Section VI} concludes our study and outlines potential future work.

\section{Related Work} \label{related-work}

\subsection{Commit Message Generation}

% Several studies have done CMG, but their approach had limitations.
Over the past few years, several studies have aimed to improve the state-of-the-art in CMG by exploring various methods to represent the changes in a commit, such as the diff \cite{shi2022raceretrievalaugmentedcommitmessage}, Abstract Syntax Tree (AST) paths \cite{Dong2022FIRA:Generation,Liu2022ATOM:Ranking}, issue states \cite{Tao2021OnStudy}, and modification embedding \cite{He2023COME:Embedding}. However, these CMG methods did not account for the impact of bot-generated and uninformative CMs during training, which has been shown to render their reported performance inaccurate \cite{zhang2024automatic-cmg-crititical-review}. 
Accordingly, researchers proposed filtering the adopted CM datasets to include only good practice CMs \cite{Tao2024KADEL:Generation} when training the deep learning models.
However, these methods relied on the quality of human-written CM, which has been reported to lack the required quality \cite{li2024omg,Tao2024KADEL:Generation,Tian2022WhatMessage}, and did not align their generated CM with practitioners' expectations of a good CM \cite{li2024omg}.

% OMG studied practitioners' expectations and proposed a CMG approach that meets their expectations.
To address these shortcomings in previous CMG methods, Li et al. conducted surveys and data mining to understand practitioners' expectations for a good CM \cite{li2024omg}. They proposed an LLM-based CMG approach called OMG that uses the ReAct prompting framework \cite{Yao2022ReAct:Models} with GPT-4 to meet the identified expectations. Based on their findings, the commit diff, which was the primary artifact used in traditional CMG methods, is not enough to generate a CM that aligns with developers' expectations. 
Hence, they utilized six different contextual pieces of information about a commit, resulting in CMs that surpassed the quality of those generated by the previous state-of-the-art, FIRA \cite{Dong2022FIRA:Generation}, as determined through human evaluation.
Despite achieving superior results, the authors did not consider using an LLM that addresses privacy and sustainability concerns. Instead, they employed the most advanced proprietary LLM available during the development of OMG. Our study builds upon OMG in terms of the commit context provided to the model and employing an LLM. However, we seek a different goal. Our objective is to generate high-quality CMs without relying on state-of-the-art proprietary LLMs.

\subsection{OLLMs in Software Engineering}
Researchers have investigated the effectiveness of OLLMs in solving various software engineering tasks and compared them with proprietary LLMs, yielding both positive and negative results across different areas \cite{hou2024largelanguagemodelssoftware}.

Yin et al. evaluated OLLMs in identifying software vulnerability. They reported that while OLLMs demonstrate some capability in specific areas, they still require further improvement to be truly effective in addressing software vulnerability-related tasks \cite{yin2024multitaskbasedevaluationopensourcellm}.
Pan et al. investigated the effectiveness of LLMs in code translation \cite{pan2024codetranslation}. Among the evaluated models, including OLLMs and GPT-4, the best performing OLLM, StarCoder, 
% \cite{li2023starcodersourceyou}, 
achieved a 14.5\% successful translation rate compared to 47.3\% by GPT-4. 

On the other hand, OLLMs have demonstrated comparable results with GPT-4 in certain areas. 
In a study by Liu et al, Vicuna 13B 
% \cite{vicuna2023} 
equipped with the author's proposed online log analysis approach, LogPrompt, showed comparable performance with GPT-4 \cite{liu2024loganalysis}.
Zhong and Wong \cite{zhong2024chatgptreplacestackoverflowstudy} inspected the reliability and robustness of the LLM-generated code. They found that although Meta Llama2 
% \cite{touvron2023llama2openfoundation} 
achieved a low API misuse rate, its compilation rate was reported significantly lower than the other models in the study. However, instruction-tuned Deepseek-Coder 6.7B 
% \cite{deepseekcoder2024} 
achieved comparable results in terms of the balance between compilation rate and API misuse rate. 
Munley et al. explored various LLMs' capabilities in generation of a validation and verification test suite for high-performance computing compilers from a standard specification. Their results indicated the instruction-tuned Deepseek-Coder 33B 
% \cite{deepseekcoder2024} 
produced the most passing tests followed by GPT-4-Turbo \cite{munley2024llm4vvdevelopingllmdriventestsuite}. 

Given the varied performance of OLLMs across different software engineering tasks, it is essential to investigate their ability to generate CMs. Our study aims to fill this gap in existing research through investigating the potential of OLLMs in producing high-quality CMs that meet practitioners' expectations.

%old: Given the varied performance of OLLMs across different software engineering tasks, their ability to generate CMs that matches or exceeds state-of-the-art quality is to be found out. Therefore, our study is the first to investigate the potential of  OLLMs in producing high-quality CMs that meet practitioners' expectations \cite{li2024omg}.

% Few studies have proposed cost-effective LLM-based solutions for software engineering tasks. Xia et al. questioned the necessity of using complex autonomous software agents to solve software development problems \cite{xia2024agentlessdemystifyingllmbasedsoftware}. They introduced an AGENTLESS approach using GPT4-o \cite{HelloOpenAI}, which outperformed other proprietary LLM-based alternatives on the SWE-BENCH Lite benchmark, achieving superior results at a lower cost. Our study differs from these as we focus on the CMG task, aiming to achieve high-quality commit messages using OLLMs.

\section{Methodology} \label{method}

\useunder{\uline}{\ul}{}

\begin{figure*}[tp!]
  \centering
  \includegraphics[width=0.9\textwidth]{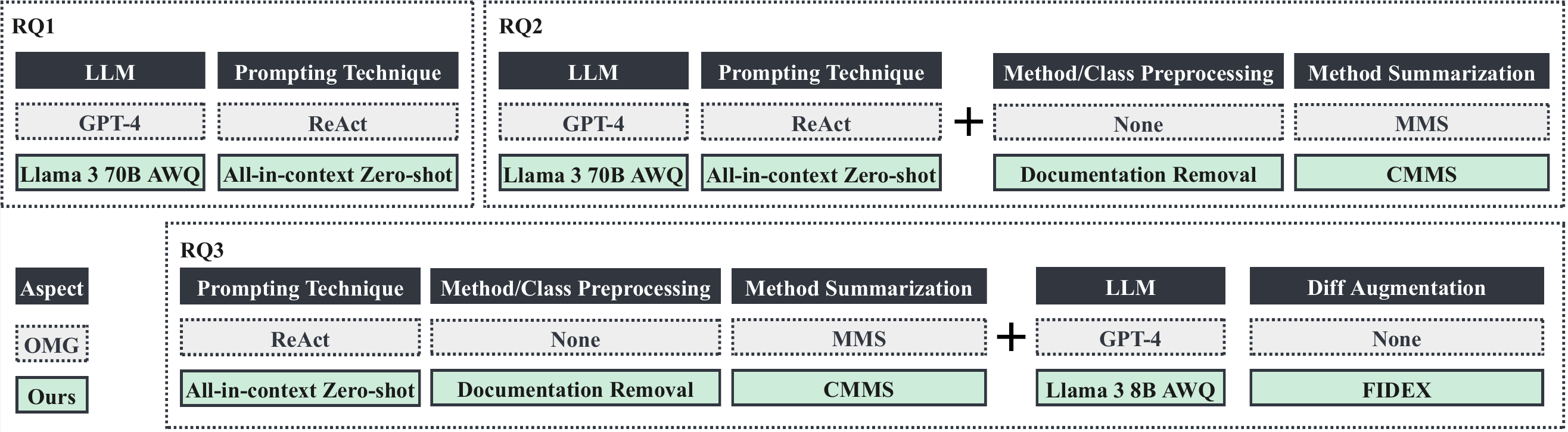}
  \caption{Overall Methodology. Modified aspects of OMG are represented using \includegraphics[width=1.35em]{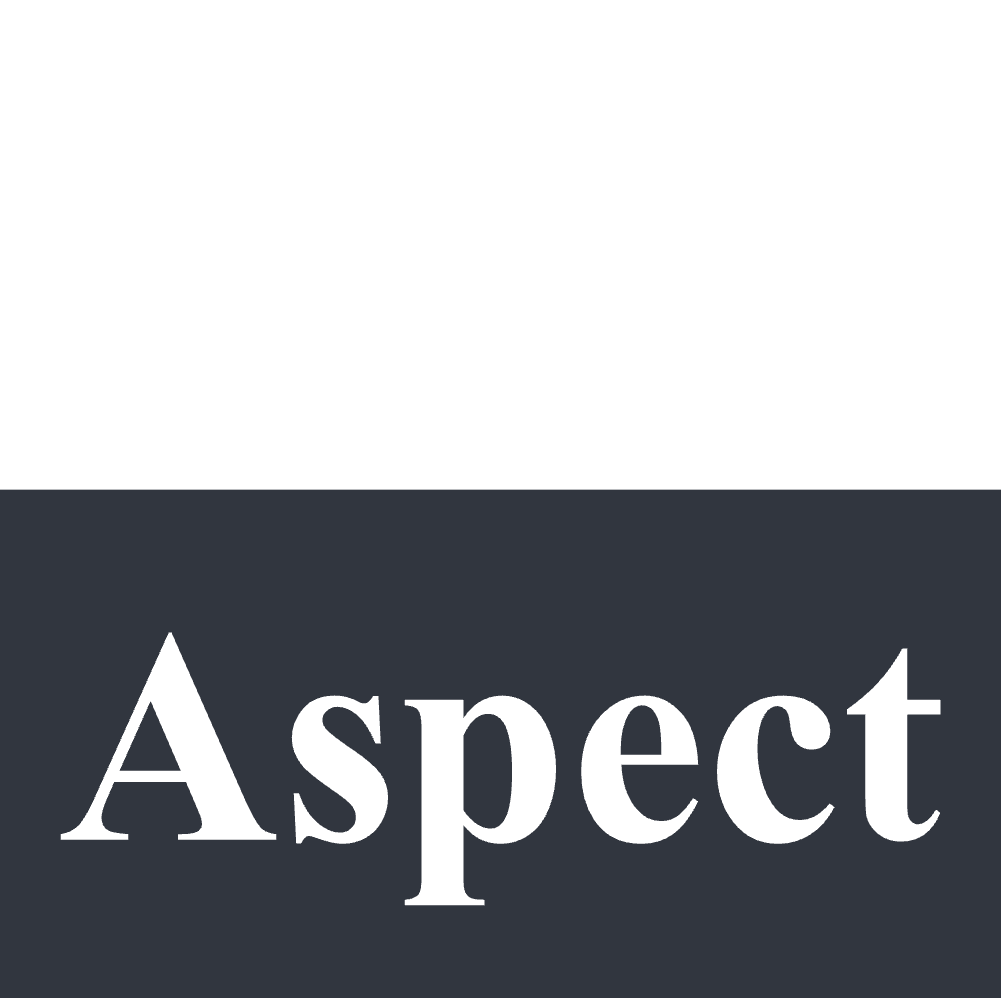}. For each changed aspect, 
    \includegraphics[width=1.35em]{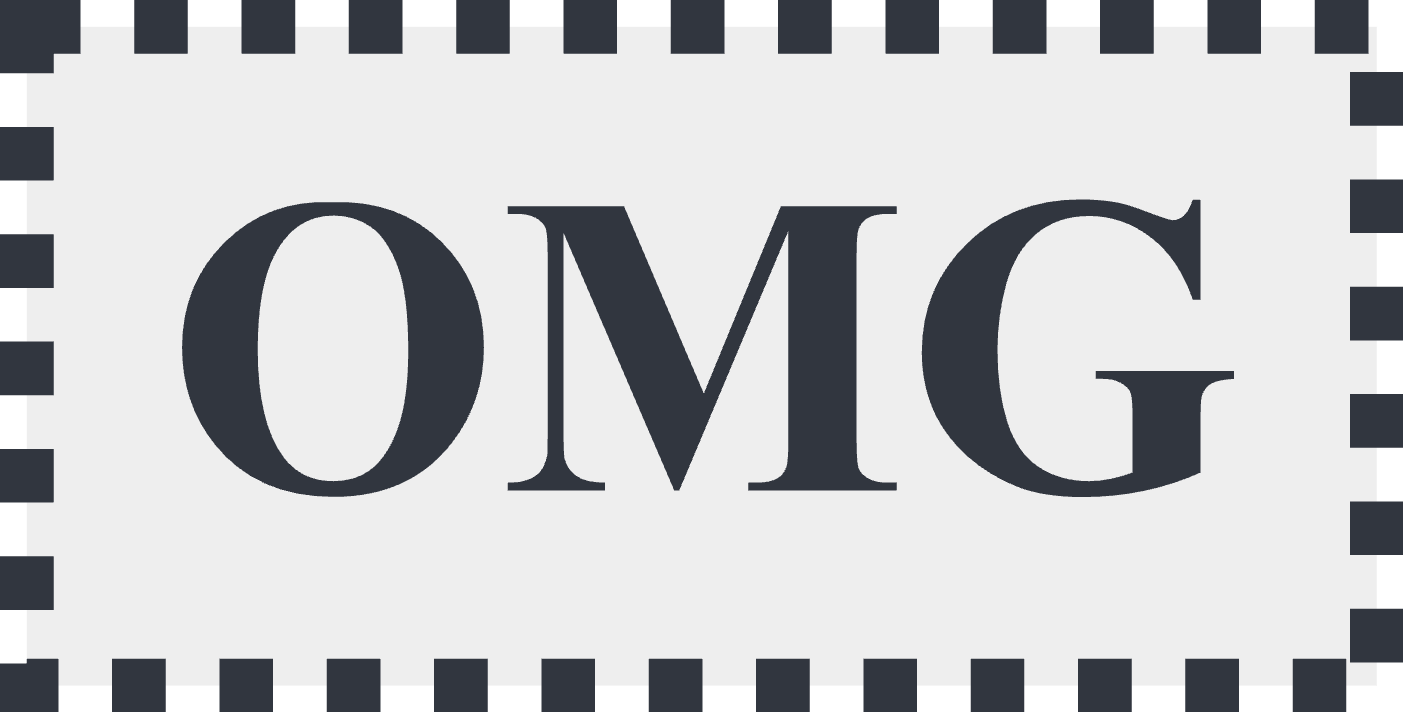} shows the value for OMG, while 
    \includegraphics[width=1.35em]{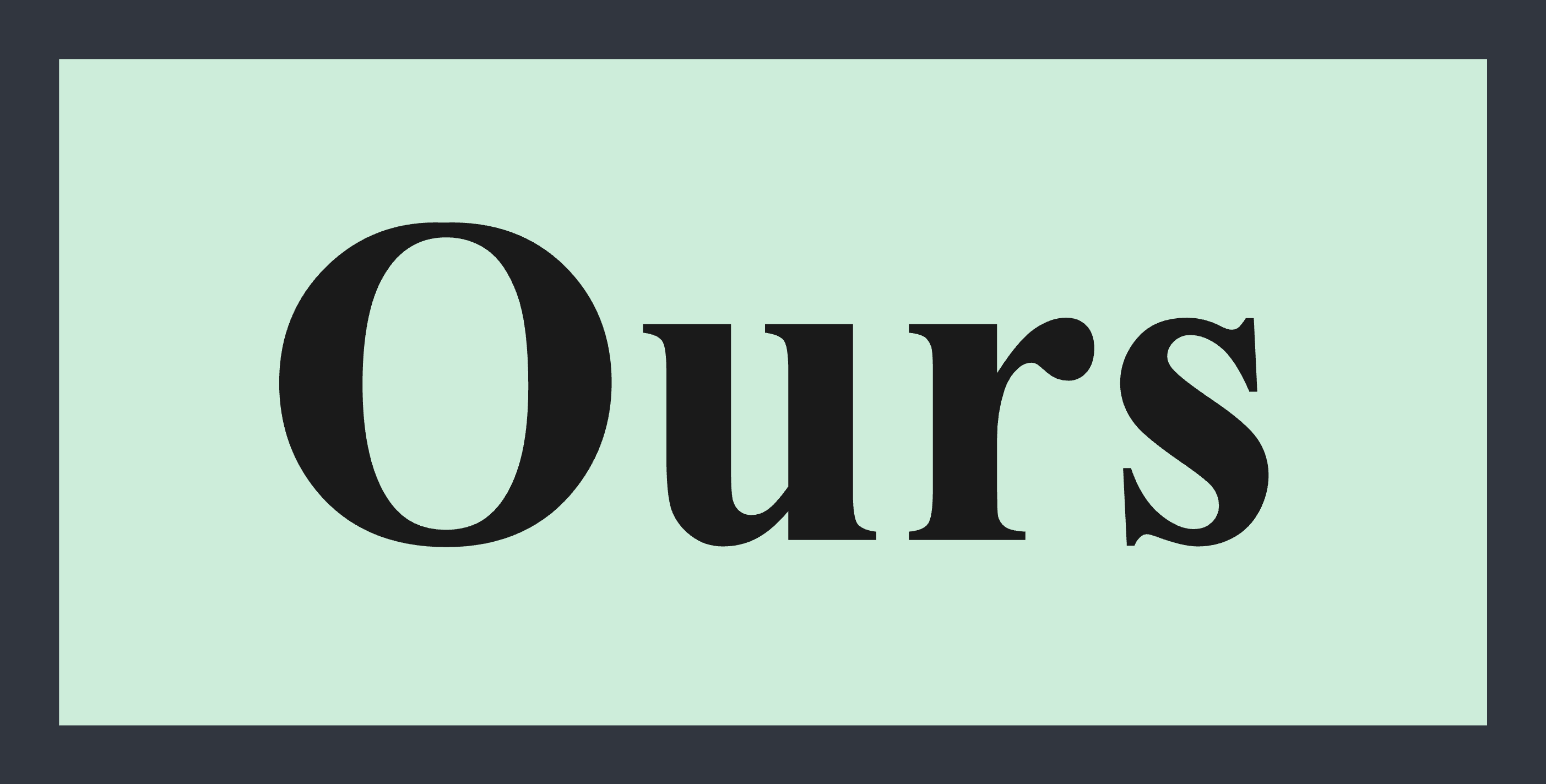} represents ours.
}
  \label{fig:overall-approach}
\end{figure*}

In this section, we present our methodology in answering our research questions.
%First, we introduce the commits dataset used and the datasets employed to validate the candidate models for producing \textit{LLM-derived commit context}. Next, we describe the experimental setup for running OLLMs for inference, including the adopted quantization technique for some candidate OLLMs in answering RQ1. We then present the experiments and context enhancements that answer all our research questions. 
%\hl{by looking at the figure I cannot understand what is has been changed/diff with OMG}
Figure \ref{fig:overall-approach} presents the changes we made to various aspects of OMG in answering each research question.
% We then explain our approach for selecting candidate OLLMs and SLMs for RQ1 and RQ3, and how we chose the final OLLM and SLM for the remaining experiments. Following this, we present the different prompting techniques tested on the selected OLLM. We also outline our survey design and recruitment strategy for our three practitioner surveys. Finally, we detail our methodology for enhancing the \textit{LLM-derived commit context} to answer RQ2 and our augmentation methods for the raw diff as part of answering RQ3.

\subsection{Base Commit Context}

In addition to the commit diff, we utilized six different contextual pieces of information about a commit that were proposed and utilized by OMG \cite{li2024omg}. Specifically, for each commit, we provided the following contextual information to the OLLM/SLM: 
1) Associated issues on the version control system or issue tracking service 
2) Associated pull requests 
3) Relative importance of changed files 
4) The software maintenance activity type of the commit 
5) Multi-Intent Method Summaries of changed methods \cite{multi-intent-2024} 
6) Summary of changed classes. 
Since the latter three components (4-6) of the commit context are generated by an LLM, we refer to them as the \textit{LLM-derived commit context}. 
% hl{\textbf{Done.} all places where we  are using the name \textit{LLM-derived commit context}, fix them by doing the the italic thing }
\label{llm-derived-context}
We used this context as the basis for our study. However, we made specific refinements when addressing our RQs, which we detail in \hyperref[method-rq2]{subsection F}.

% \hl{\textbf{Added two sentences.} why are you telling me about MMS?}
One of the contextual refinements we propose in this work is altering how method summaries were generated for the affected methods. Therefore, it is important to discuss the original approach adopted by OMG. 
OMG employs the Multi-Intent Method Summarization (\textbf{MMS}) technique proposed by Geng et al. \cite{geng2024multi-intent} to summarize the methods affected by a commit, 
% \hl{MMS is employed by OMG,}
% \hl{here is a major disconnect, why are you telling me about MMS, lets talk about this one}
% To answer RQ2, we needed to do refinemen LLm derived commit context. We specifically focused on MMS because ...(explained in details in Section Blah). 
% Multi-Intent Method Summarization (\textbf{MMS}) technique proposed by Geng et al. \cite{geng2024multi-intent}
MMS is an LLM-based method comment generation approach that uses few-shot prompting to generate the summaries for a method from five different aspects (Developer's intents) as follows:
\label{multi-intent-aspects}
1) \textbf{What} describes the functionality of a method
2) \textbf{Why} Explains the reason why a method is provided or the design rationale of the method
3) \textbf{How-to-use} Describes the usage or the expected set-up of using a method
4) \textbf{How-it-is-done} Describes the implementation details of a method
5) \textbf{Property} Asserts properties of a method including pre-conditions or post-conditions of a method.

Later in this section (See \hyperref[method-rq2]{subsection F}), we discuss how MMS is employed by OMG, why it is not fit for the CMG task, and how we can overcome its limitations.

\subsection{Datasets}

Since our work builds on OMG, our dataset should include practitioner-evaluated CMs generated by OMG. This ensures our ground truth CMs have high human evaluation scores \cite{li2024omg}. The 
dataset includes 381 commits from 32 Apache projects in Java. We used this dataset to compare our CMs generated using our approach with those generated by OMG using automated evaluation metrics and practitioner surveys. 

Additionally, we used the same dataset to evaluate the performance of our candidate models in producing \textit{LLM-derived commit context} (as discussed \hyperref[llm-derived-context]{earlier} in this section). Specifically, for the software maintenance activity type classifier, we used the same dataset of 1,151 commits in Java manually labeled with three maintenance activities (Corrective, Perfective, and Adaptive).
% by Levin et al. \cite{Levin2017BoostingChanges}. \hl{why?}
% \hl{\textbf{Done.} fix}
To evaluate the performance of candidate OLLMs/SLMs in class summarization, similar to the approach taken by Li et al. \cite{li2024omg}, due to budget constraints and the high cost of using GPT-4, we sampled 384 class-summary pairs (confidence level 95\%, margin of error 5\%) from the class summary dataset used by OMG.
% OMG utilizes the Multi-Intent Method Summarization (\textbf{MMS}) technique proposed by Geng et al. to summarize the affected methods in pre- and post-commit states. MMS is an LLM-based method comment generation approach that uses few-shot prompting to generate the summaries for a method from different aspects (Developer's intents), namely, its functionality, purpose, usage, implementation details, and the pre-conditions or post-conditions of the method.
Lastly, to examine our candidate OLLMs/SLMs in generating method summaries, we used the test set utilized by OMG. All these datasets are provided in the supplementary \cite{icse25replication}.

%provided by Geng et al. \cite{geng2024multi-intent} and

% \hl{since you are using the OMG data in 2 cases, why not for that one case? and we should only talk about that specific case.}

\subsection{Evaluation Metrics}

\subsubsection{Automated Metrics}
OMG-generated CMs have been evaluated by practitioners and achieved high human evaluation scores \cite{li2024omg}. Therefore, we postulated that if we make our CMs similar to those generated by OMG, they would achieve acceptable results when evaluated by practitioners. This approach allowed us to use the similarity to OMG-generated CMs as an initial quality assurance measure before conducting human evaluations.
% , ensuring that human evaluation is performed only once the generated CMs exhibit high similarity to CMs that were previously rated highly by practitioners.
Accordingly, we used standard evaluation metrics that are used to compare CMs with state-of-the-art machine-generated CMs \cite{Dong2022FIRA:Generation,He2023COME:Embedding, li2024omg}. Specifically, we used BLEU, 
% \cite{sacrebleu}, 
METEOR, 
% \cite{banerjee2005meteor}, 
and ROUGE-L 
% \cite{lin2004rouge} 
to measure the similarity between the CMs generated by an OLLM and SLM with those generated by OMG. 
Additionally, following the common practices in CMG research \cite{li2024omg}, we reported automated metrics by comparing our CMs with human-written CMs.

\subsubsection{Human Metrics}
In order to evaluate our CMs by practitioners, we opted for using the four human evaluation metrics proposed by Li et al. in our surveys.
% We used the four human evaluation metrics proposed by Li et al. 
These metrics were developed through careful study of CMG literature and discussions among researchers \cite{li2024omg}. The metrics are: 1) \textbf{Rationality}, which assesses whether a CM provides a logical explanation for the code change and identifies the software maintenance activity type. 2) \textbf{Comprehensiveness}, which evaluates whether the message summarizes what has been changed and includes relevant important details. 3) \textbf{Conciseness}, which measures the brevity of a CM. 4) \textbf{Expressiveness}, which examines the grammatical correctness and fluency of the CM. 
% \hl{\textbf{Added one sentence to the beginning of the para.} while conducting the survey we asked the participants to evaluate out CM using these four metrics, .......}

\subsection{Experimental Setup}

We utilized a Linux server equipped with an NVIDIA A6000 GPU with 48GB of VRAM to run the OLLM inference engine. 
The 48GB of GPU VRAM limited our experiments to OLLMs with up to 20 billion trained parameters in full precision or up to a 4-bit quantized 70 billion parameter OLLM 
% \cite{XiongjieDai/GPU-Benchmarks-on-LLM-Inference:Inference, LLMFernando}. 
\cite{DaiGPU-Benchmarks-on-LLM-Inference}. 
% \cite{LLMFernando}. 
Therefore, we had to select a quantization method to quantize OLLMs with more than 20 billion trained parameters. Among the state-of-the-art quantization methods,
% \cite{Lin2023AWQ:Acceleration, frantar2023gptqaccurateposttrainingquantization}, 
we chose Activation-aware Weight Quantization (\textbf{AWQ}) due to its minimal impact on model's perplexity and the inference speedup it provides for the quantized LLM \cite{Lin2023AWQ:Acceleration}. For the remainder of this paper, a quantized model refers to an OLLM that has been quantized to 4-bit using the AWQ technique.

% \hl{\textbf{Simplified.} start by teaching us what is intfenerence engine, why its used and then tell}
To efficiently manage and execute an OLLM, we chose VLLM \cite{vllm2023} due to its efficient memory management, compatibility with our server, and ease of deploying a wide range of OLLMs.
% \cite{wolf2020huggingfacestransformersstateoftheartnatural}. 
VLLM provides an OpenAI-compatible API, facilitating its use with LLM app developments such as Langchain \cite{LangChain}, 
which we used to develop the LLM agents utilized in this study. However, at the time of our experiments, VLLM lacked the embedding inference capability. Consequently, we could not use the original class summarization approach employed by Li et al., which relied on embedding the changed classes and using retrieval-based question answering to summarize them. Instead, we utilized zero-shot prompting to summarize the affected classes.
Lastly, to ensure consistent output, we set the temperature to 0 when using the deployed OLLM for inference, similar to previous work 
% \cite{xu2024unilog, ren2024reflectioncoderlearningreflectionsequence, li2024omg}.
\cite{li2024omg}.
%adopted greedy decoding strategy
\subsection{Survey}

% \hl{Look for anything about the correlation analysis and remove it}

Overall, we conducted three practitioner surveys to answer our research questions. %Following, we review our survey methodology, including commit sampling, participant recruitment, and survey design.

\subsubsection*{\textbf{Commits Sampling}}
%Given the similarity of our surveys to the one conducted by Li et al., 

We adopted a similar approach as Li et al. while conducting our surveys to keep the workload manageable for participants. Specifically, evaluating CMs for the entire commit dataset, which involves assessing 762 candidate CMs for 381 commits from four perspectives, would be impractical for participants. Therefore, we randomly sampled 15 commits for each survey, resulting in 30 commit messages (15 CMs generated using our approach and 15 CMs generated using the approach we are comparing with) for participants to compare and evaluate. This is the same total number of CMs that were evaluated by the survey participants in the evaluation of OMG, ensuring a fair workload to achieve a high completion rate \cite{li2024omg}.
% \cite{smith2013developer-participation-rate}. 

\subsubsection*{\textbf{Participant Recruitment}}
% To ensure data quality and avoid the challenges associated with recruiting survey participants through social media \cite{Pozzar2020Threats}, 
We adopted the snowball sampling approach to recruit the participants for our surveys \cite{wohlin2014snowball}. Specifically, we began by distributing the survey to our industry contacts with at least two years experience in Java development and sent periodic reminders to encourage participation. Additionally, we asked them to share the survey with other developers who have relevant programming backgrounds. This approach ensured that participants possessed the necessary knowledge to accurately assess the CMs generated for Java projects.

\subsubsection*{\textbf{Survey Design}}

All three surveys were designed to comparatively evaluate CMs written for a sample of 15 commits. For the first and last surveys, we presented two candidate CMs for each commit (CM \#1 and CM \#2). 
%In 8 out of 15 questions, the OLLM/SLM-generated CM was CM \#1, and in the remaining 7 questions, CM \#1 was generated by OMG. 
We randomly shuffled the questions to eliminate any bias towards an option due to an observable pattern in the candidate CMs. 
The survey was hosted on QuestionPro \cite{QuestionPro}, and participants were provided with definitions of the human evaluation metrics. Following the definitions, we presented the commits with all the commit context along with the two candidate CMs.
For each human evaluation metric, we asked the participants to select their preferred CM or choose ``Identical'' if there was no clear preference. The Identical option was provided to see if the CMs generated by the OLLM/SLM are ``comparable'' to those produced by OMG. Additionally, for each commit, we asked the participants to select their overall preferred CM.

The second survey was designed differently, as its purpose was to validate the comprehensiveness of the CMs after making LLM-derived commit refinements. For each commit, we presented the CMs before and after the refinements. Similar to the other surveys, we randomly shuffled the questions. For each commit, we asked the participants to choose a candidate CM that is more comprehensive. We did not provide an ``Identical'' option for this survey, as our goal was to assess whether the enhanced context produced more comprehensive CMs, not just comparable ones.

\subsection{Answering RQs}
In the following subsections, we detail the steps taken to address each research question.

\subsection*{\textbf{RQ1.} Can an OLLM generate CMs comparable to a state-of-the-art LLM (GPT-4)?}

\subsubsection*{\textbf{OLLM Selection}}

To answer RQ1, we needed to compile a list of candidate models. Given the reliance of OMG on code summarization and the necessity of code understanding to generate high-quality CMs, candidate models had to demonstrate strong performance on code-related tasks by scoring high on relevant benchmarks. %Additionally, we had to consider our hardware limitations. Taking these factors into account, 
We selected four OLLMs as candidates to answer RQ1. All selected models held leading rankings in the EvalPlus leaderboard \cite{Liu2023IsGeneration} and the Big Code Models leaderboard \cite{bigcode-evaluation-harness} at the time of compiling the candidate list. 
EvalPlus is a code synthesis evaluation framework designed to benchmark the functional correctness of LLM-synthesized code. The Big Code Models leaderboard evaluates the performance of base multilingual code generation models on the HumanEval benchmark 
% \cite{chen2021humaneval} 
and MultiPL-E. 
% \cite{cassano2022multipl-E}. 
These benchmarks are commonly referenced by researchers when selecting LLMs \cite{du2024}.
We chose a quantized instruction-tuned Llama3 70B \cite{llama3} and a quantized instruction-tuned DeepSeek-Coder 33B \cite{deepseekcoder2024}. 
We selected instruction-tuned Llama3 70B since
%Although the instruction-tuned Llama3 70B was not listed on either leaderboard at the time we compiled our candidate OLLMs, 
its score on the HumanEval benchmark was comparable to the top-3 models on the Big Code Models leaderboard \cite{llama3}. We also selected the instruction-tuned DeepSeek-Coder 33B because it was ranked as one of the top models in the EvalPlus benchmark.

In addition, we used the AutoAWQ library to quantize CodeFuse-DeepSeek-33B and OpenCodeInterpreter-DS-33B \cite{zheng2024opencodeinterpreterintegratingcodegeneration}, as AWQ quantized versions of these models were not available on Huggingface. 
When sorting the models on the Big Code Models leaderboard by their performance in generating correct Java code, CodeFuse-DeepSeek-33B and OpenCodeInterpreter-DS-33B were ranked 
among the top models.
% 2nd and 3rd, respectively.
% The Huggingface URLs for all the candidate OLLMs are provided in the supplementary \cite{icse25replication}. \hl{same attacking point, why not the top ones?}

In order to ensure the quality of the \textit{LLM-derived commit context}, similar to Li et al. \cite{li2024omg}, we evaluated the candidate OLLMs performance on class summary generation, MMS, and classifying software maintenance activity types. Based on the evaluation results, we selected the top performing model, quantized instruction-tuned Llama3 70B, for our experiments to answer RQ1 and RQ2.

%\textit{quantized instruction-tuned Llama3 70B}  

% \hl{\textbf{Updated here. }update intro or here to  address the inconsistency in the para starting with secondly}
\subsubsection*{\textbf{Prompting Method}}
\label{rq1-prompting}

In answering RQ1, our goal was to determine the feasibility of replacing GPT-4 with an OLLM in the original implementation of OMG. Therefore, although the adopted prompting method, ReAct \cite{Yao2022ReAct:Models}, does not align with one of the primary objectives of our study, sustainability of the CMG approach, we chose to begin our experiments with this prompting method. This decision was made to limit the changes to the adopted LLM.
% We initially experimented with the ReAct prompting framework \cite{Yao2022ReAct:Models} that was adopted by OMG. 
However, we observed poor automated scores when comparing the CMs generated using ReAct by our selected OLLM to those generated by OMG. This led us to question the capability of the OLLM in generating useful thoughts to reason about each contextual piece of information about a commit, as noted by other researchers \cite{xing2024understandingweaknesslargelanguage}. 

% Since in RQ1 our goal is to conduct a fair comparison by isolating the variation to only changing the adopted LLM from GPT-4 to an OLLM, 
To address the poor performance, we avoided using more advanced prompting techniques to ensure the results did not stem from improvements in the adopted prompting approach.
Hence, we utilized the simplest prompting method, zero-shot prompting \cite{liu2021pretrainpromptpredictsystematic}. Instead of expecting the OLLM to request each piece of commit context as in ReAct, we provided all the available context about a commit along with CMG instructions in one prompt. This change in prompting technique increased the BLEU score achieved by the OLLM by 118\%. 
The high automated scores (METEOR and ROUGE-L above 30) from zero-shot prompting suggested that the generated CMs resembled those produced by OMG. To verify their quality, we conducted a survey with practitioners to compare the CMs from the selected OLLM with those from OMG using human metrics. The survey results are discussed in \hyperref[results]{Section IV}.

The survey verified our assumption. However, practitioners preferred OMG-generated CMs for their comprehensiveness, which led to our second research question.

%The high automated scores achieved by the zero-shot prompting (METEOR and ROUGE-L higher than 30) suggested that the generated CMs were similar to those produced by OMG. To validate if that means they match the quality of OMG-generated CMs, we conducted a survey with practitioners to comparatively evaluate the CMs produced by the selected OLLM with those generated by OMG in terms of human metrics. We discuss the survey results in Section IV. %Briefly, although the survey verified our assumption, practitioners preferred OMG-generated CMs in terms of comprehensiveness.

\subsection*{\textbf{RQ2.} How can we bridge the comprehensiveness gap between the CMs produced by an OLLM and the CMs generated by a state-of-the-art LLM (GPT-4)?}
\label{method-rq2}
% \subsubsection{\textit{LLM-derived commit context} Refinements}

To answer RQ2, 
% which aims to bridge the identified comprehensiveness gap between the CMs generated by OMG and those generated by the selected OLLM, 
we initially expanded our zero-shot prompt to ask the OLLM to comprehensively detail all the changes that occurred in the commit without missing anything. 
However, this not only failed to increase the automated metrics but also moderately decreased them.

Therefore, since a change in the prompt did not address RQ2, we shifted our focus to the provided context. Specifically, we investigated the effectiveness of the \textit{LLM-derived commit context}, namely, the summaries of changed classes and methods, and the software maintenance activity type. 
Our CMs were structured similarly to OMG; they included a header with the software maintenance activity type and a brief subject, followed by the body. Consequently, the software maintenance activity type contributed only one word to the CM, i.e., one of refactor, fix, style, or feat \cite{li2024omg}. Therefore,
% Since the software maintenance activity type classifier contributes only one word to the CM
we concentrated on the class and method summaries, positing that improvements in these should help the OLLM write more comprehensive CMs.

According to the literature, mismatched code comments cause the models to produce summaries that reflect the comments rather than the actual code functionality \cite{song2022}.
% \hl{thsi following sentence is unnecessary}
% Additionally, since the summaries of the classes and methods are provided to the LLM to capture the changes to the functionality at a higher granularity than the provided diff, changes to the code documentation should not result in a different summary.
Hence, we added a preprocessing step to the class and method bodies passed to the OLLM for summarization by removing all the documentation (comments and Javadocs). 
% \hl{repeated}
% This ensures that the generated summaries rely solely on the code lines, preventing any flaws in the comments from negatively impacting the summaries.

% The rationale behind this additional preprocessing is that the summaries of these program entities are provided to the model to capture any changes to the functionality at a higher granularity than the provided diff.

% Since changes in documentation do not affect the functionality of classes or methods, we removed them to prevent any errors by the OLLM in generating the summaries. 

Furthermore, we changed the way the MMS was done for modified methods (methods that exist in both pre- and post-commit states). Originally, OMG generated these summaries separately for the pre-commit and post-commit bodies of affected methods and provided these as the summaries of the affected methods. However, we hypothesize this approach is not suitable for a code change-based task like CMG, as it does not ask the OLLM to generate summaries based on how the changes in the diff impact each of \hyperref[multi-intent-aspects]{the five different aspects} of affected methods. Instead, this approach assumes LLM's ability in generating different summaries with slight changes to the method bodies. 

In order to validate our assumption, we randomly sampled 55 commits from the 280 commits in which the changes affected a method (confidence level 90\%, margin of error 10\%). Two authors independently compared the generated summaries for the pre- and post-commit bodies of the affected methods to see if the changes are correctly captured (yes/no) in the summaries for any of the five aspects. 
% \hl{\textbf{Done.} fix english for the following sentence}
We observed that in 96\% of the commits (inter-rater agreement of 88\%), there was no conceptual difference in the generated method aspects (recall \hyperref[multi-intent-aspects]{the five aspects} of a method that MMS provides) by the OLLM for the method bodies before and after the commit. This confirmed our hypothesis.

To address the identified shortcoming of MMS, generating semantically identical  summaries for affected methods before and after the commit, we introduce \textbf{C}hange-based \textbf{M}ulti-Intent \textbf{M}ethod \textbf{S}ummarization (\textbf{CMMS}). Specifically, for a modified method, we first generated the pre-commit multi-intent summaries using the original approach proposed by Geng et al. \cite{multi-intent-2024}. Next, we passed the pre-commit summary along with a list of changes to the pre-commit method body, produced by a Python script that parses the method body before and after the commit. We then asked the OLLM to explain how each method aspect of the pre-commit method body would be affected by these upcoming changes.
For instance, changes to a method's input arguments affect the \textit{How-to-use} aspect of it.

% \hl{Write something about we conducted an ablation study...}
We conducted an ablation study to examine the effectiveness of each refinement (documentation removal and CMMS) separately. Both refinements improved all automated metrics, using OMG as the reference. We report the automated evaluation results in \hyperref[results]{Section IV}. 
Since the OMG-generated CMs were perceived as more comprehensive by the participants in our first survey, this increased similarity to OMG-generated CMs led us to infer that the produced CMs had become more comprehensive. To validate this assumption, we conducted a second survey and asked participants to compare the CMs produced with the old commit context to those resulting from the enhanced commit context. The survey results verified our hypothesis. The survey results are reported in \hyperref[results]{Section IV}.

\subsection*{\textbf{RQ3.} Can a smaller OLLM (SLM) produce CMs comparable to a state-of-the-art LLM (GPT-4)?}

% \subsection{CMG With SLM}

\subsubsection*{\textbf{SLM Selection}}
\label{slm-selection}
To maintain a fair comparison between the performance of a selected SLM and the quantized Llama3 70B, we only considered the quantized versions of any SLM. Given the acceptable performance of the quantized Llama3 70B, we included the lighter version of the Llama3 family, the quantized instruction-tuned Llama3 8B, as one of the candidate SLMs. Additionally, we considered the quantized versions of conversation-tuned CodeQwen1.5 7B \cite{codeqwen1.5} and instruction-tuned Mistral v0.3 7B \cite{jiang2023mistral7b}. At the time of gathering the candidate SLMs, these three models had the highest rankings among OLLMs with under 10 billion trained parameters in the RepoQA benchmark, which evaluates LLMs' capability in long-context code understanding tasks \cite{liu2024repoqaevaluatinglongcontext}.

Similar to our approach for OLLM selection in RQ1, we evaluated the candidate SLMs on class summarization, MMS, and software maintenance activity classification tasks.
Initially, we selected the conversation-tuned CodeQwen1.5 7B due to its superior performance in generating the \textit{LLM-derived commit context}. 
However, after observing its output in an initial CMG experiment, we noticed that the model often repeated the same sentences to fill the allowed maximum tokens. While we could have used a repetition or frequency penalty to mitigate this issue, best practice from the relevant research dictates that the model should not get penalized for reusing tokens \cite{galindo2023}.
% \hl{\textbf{Done.} Summarize and fix writing}
Hence, we used the second-best model, quantized instruction-tuned Llama3 8B, for the remaining experiments without any penalties. 
% This also ensures the result do not stem from the potential changes in the SLM's behaior as the result of penalizing it for reusing tokens.

% \hl{\textbf{Did not add this due to MAX citation reached.} This also ensures the result stems from the context but not due to hyperparamter change.}

%This decision was made to ensure that the model does not get penalized for repeating any tokens, similar to the approach taken by other researchers \cite{galindo2023}. 

%Avoiding these penalties helps to prevent any negative impact on the human evaluation metrics achieved by the resulting CMs.

% Hence, we chose \textbf{quantized instruction-tuned Llama3 8B} for the remaining experiments.

\subsubsection*{\textbf{CMG Experiments}}

Given the improved comprehensiveness achieved in RQ2 and the overall good performance of our zero-shot prompting in RQ1, we experimented with the selected SLM under zero-shot prompting using the enhanced \textit{LLM-derived commit context}.
However, the automated scores were considerably lower than those achieved by the selected OLLM. Given the identical commit context, we posit that the lower automated metrics are due to the model's inability to correctly and comprehensively understand the changes introduced by the commit. 
Dong et al. found a similar issue in learning-based CMG approaches by observing that their poor quality lies in strong attention weights for marks (+/-/white spaces line prefixes) in a diff \cite{dong2023}. Nonetheless, they did not evaluate LLMs' understanding of a git diff and its impact on the generated CM. 
Accordingly, to validate our hypothesis, we randomly sampled 38 commits and asked the selected SLM to explain the changes in the diff of those commits. Next, two authors independently evaluated the SLM's answers and marked each answer as correct if all the changes in the diff were correctly covered, otherwise incorrect. We found out that in 82\% of the cases, the LLM's response was not correct. The inter-rater agreement for this analysis study was 100\%.

\subsubsection*{\textbf{FIDEX}}

In light of these findings, we realized the necessity of a diff augmentation approach that helps the SLM correctly understand the changes in a diff. This augmentation approach should address specific considerations. Firstly, it should be able to accurately comprehend all the changes that occurred in the diff without any errors (\textit{C1}).
% Since we observed the SLM's inaccuracy in explaining the changes in a diff, we opted to propose a deterministic error-free approach to address C1.
Additionally, research suggests that explicit prompting boost the performance of LLMs in inferential reasoning tasks \cite{moghaddam2023boostingtheoryofmindperformancelarge}, such as ours, which is inferring the changes in a diff during CMG. Hence, the diff augmentation approach should be able to remove this inference step for the SLM in the CMG task by explicitly stating the differences between the old and new versions (\textit{C2}). Furthermore, relevant studies in CMG suggest that providing fine-grained details of the changes in a diff can boost the performance \cite{Wang2024Multi-grainedGeneration}. Therefore, the diff augmentation approach should provide details about the differences (\textit{C3}). 

\begin{figure}[b]
  \centering
  \includegraphics[width=0.9\linewidth]{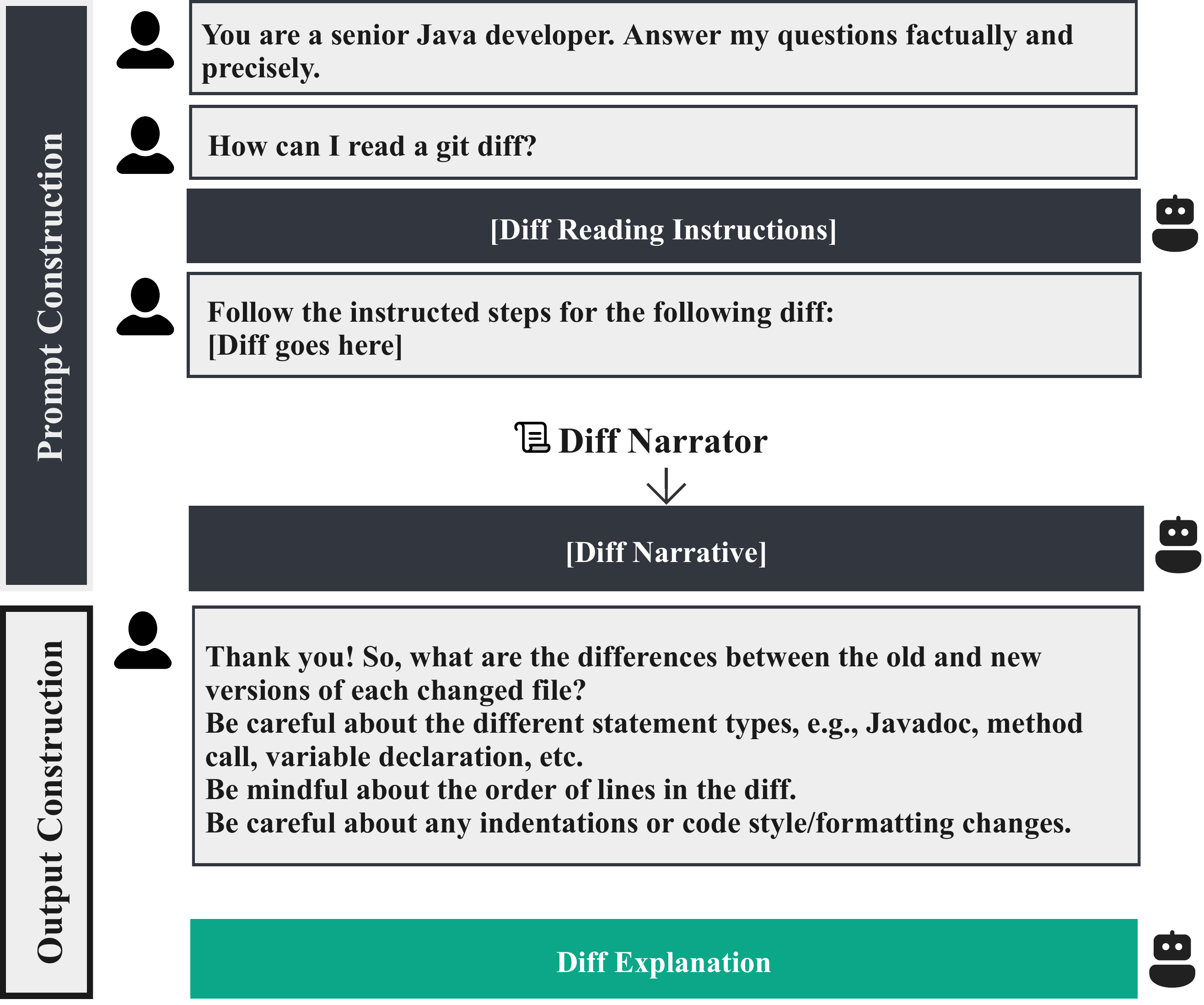}
  \caption{Fine-grained Interactive Diff Explainer. Values wrapped in square brackets are hard-coded. 
  Except the last message, other messages are fed into the LLM's memory.
  }
  \label{fig:FIDEX}
\end{figure}

% \subsubsection*{\textbf{FIDEX}}
% Cite Multi-grained contextual code representation learning for commit message generation as a motivation to represent the changes in finer details

% Since Diff Narrator only satisfies C1, we need to extend it further to satisfy the remaining considerations, C2 and C3.
% We did not find a similar approach in the literature that focuses on descriptively highlighting the differences between the old and new versions of affected Java files. 
We did not find a similar approach that addresses these considerations.%, which is highlighting the differences between the old and new versions of affected Java files in detail.
Therefore, we propose \textbf{F}ine-grained \textbf{I}nteractive \textbf{D}iff \textbf{EX}plainer \textbf{(FIDEX)}, a hybrid LLM-based approach to produce a detailed explanation of a diff, highlighting all differences between the pre- and post-commit versions of affected Java files. FIDEX is a hybrid approach since it leverages a deterministic solution to understand the changes in a diff (C1) and uses an LLM based solution while explaining the differences between the old and new versions (C2 \& C3). 

Specifically, since we observed the SLM's inaccuracy in understanding the changes in a diff, we devised a deterministic solution to minimize the hallucinations by FIDEX in understanding the changes (C1).
Particularly, we developed a Python script named \textit{Diff Narrator}. Given a commit diff, Diff Narrator outputs a \textit{Diff Narrative}, which is a numbered list of \textit{Change Items}. Change Items are basic units of changes in a diff and can be either of the following cases: 
1) \textit{Addition Chunk} Consecutive lines with `+' mark 
2) \textit{Removal Chunk} Consecutive lines with `-' mark
3) \textit{Replacement Chunk} A Removal chunk immediately followed by an Addition chunk.

% We used the Diff Narrative to prevent any hallucination caused by inaccurate understanding of the diff by the LLM. 
%\hl{\textbf{Done.} no reference ot his figure in the text, also why did we do this? it's not clear, currently it feels like we tried this and it worked. rewrite. I think break into subsubsections will help}

To explain the differences between the old and the new version of affected Java files (C2 \& C3), FIDEX prompts an LLM. We adopted the role-playing prompting technique in FIDEX, which has been shown to outperform zero-shot prompting in several reasoning benchmarks \cite{kong2024betterzeroshotreasoningroleplay}. Specifically, FIDEX consists of two phases: Prompt Construction and Output Construction. 
Figure \ref{fig:FIDEX} illustrates the FIDEX approach as a conversation between the user and an LLM.

In the Prompt Construction phase, we prepare the role prompts to be passed to the LLM. The prompts start with a Role-Setting prompt \cite{kong2024betterzeroshotreasoningroleplay}, where we define the LLM as a senior Java developer and instruct it to answer questions factually and precisely. Following this, we include two rounds of User-LLM interactions. In the first round, the user asks the LLM to provide instructions on how to read a diff. The LLM’s response includes predefined instructions for reading a diff. In the second round, the user asks the LLM to follow these instructions and describe all the changes in the input diff. To generate the LLM’s response for this prompt, we use the Diff Narrative produced by Diff Narrator.

\begin{figure}[t]
  \centering
  \includegraphics[width=0.8\linewidth]{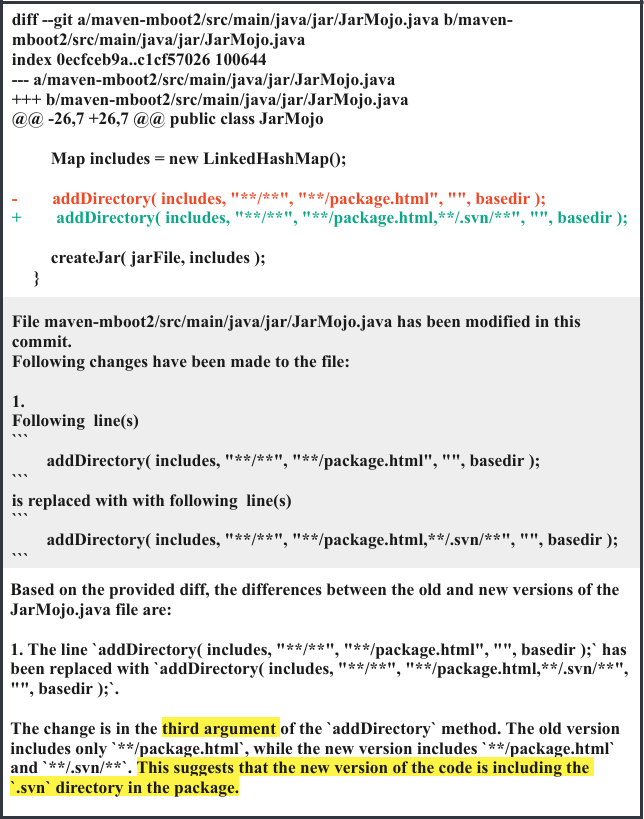}
  \caption{Example of a diff, its Diff Narrative, and its FIDEX-generated diff explanation using quantized Llama3 8B (From top to bottom). The highlighted parts show how diff explanation outlines the difference between the old and new version in fine detail.}
  \label{fig:diff-desc-example}
\end{figure}

Lastly, the Output Construction stage is where the final interaction between the user and LLM takes place. The user asks the LLM to describe the differences between the old and new versions of each affected file (C2), while considering specific cautions to ensure accuracy and comprehensiveness.
The cautions are designed to warn the LLM about different fine-grained statement types (C3), such as Javadocs, method declarations, etc., to avoid confusion between documentation and code changes \cite{fan2024exploringcapabilitiesllmscode}. Additionally, the LLM is instructed to be mindful of the order of changes to highlight reordering changes and to pay attention to the sequence of lines in the diff. The final consideration is asking the LLM to differentiate between code style or formatting changes, which helps identify stylistic changes and correctly differentiate between different software maintenance activity types in the commit \cite{li2024omg}.
Figure \ref{fig:diff-desc-example} provides an example of a raw diff, the \textit{Diff Narrative} produced by the \textit{Diff Narrator} (in grey), and a diff explanation that is generated by the selected SLM through using the FIDEX.

\begin{figure}[t]
  % \centering
  \includegraphics[width=0.90\linewidth]{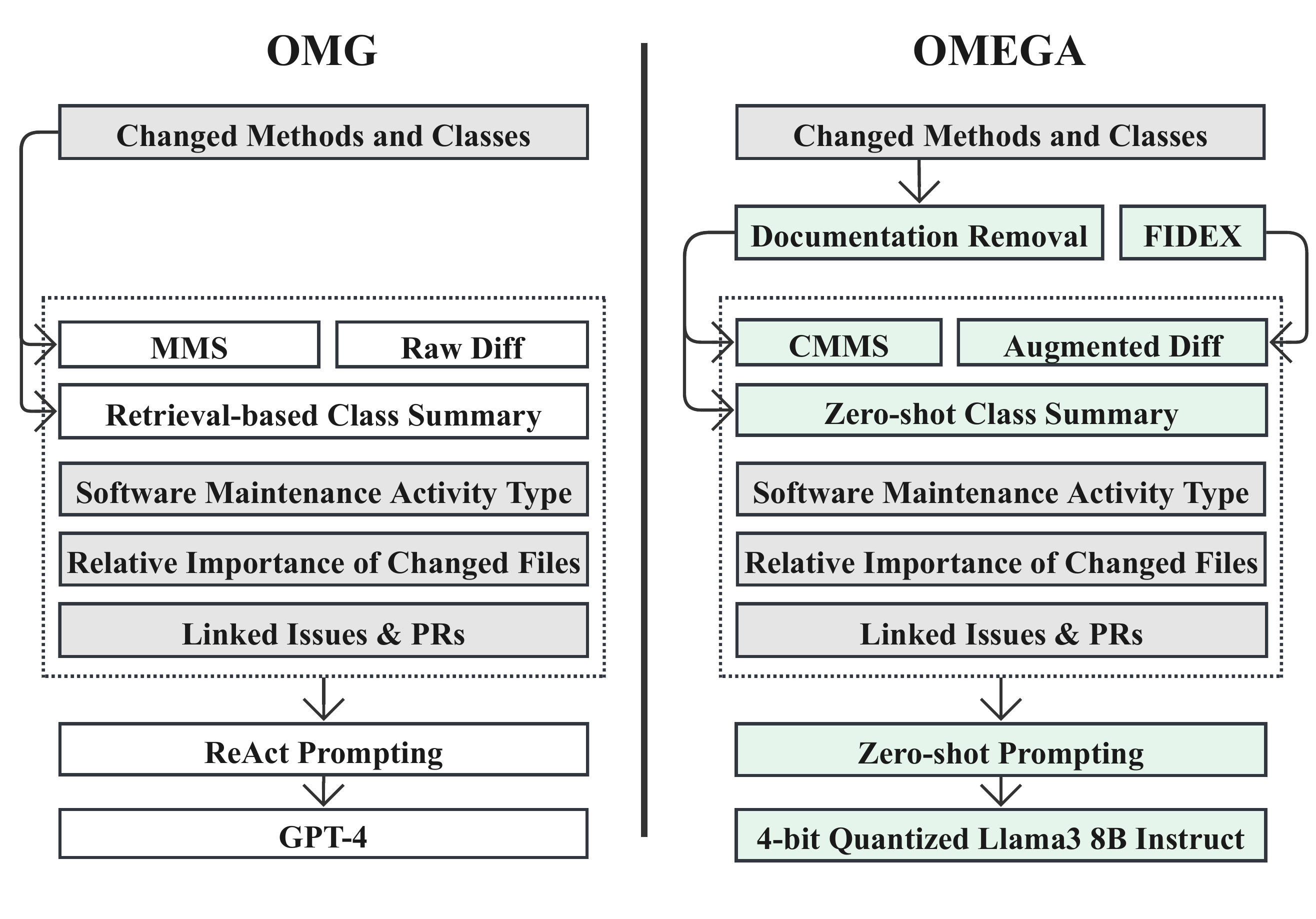}
  \caption{Differences between OMG and OMEGA}
  \label{fig:us-vs-omg}
\end{figure}

We augmented the raw diff with the FIDEX-generated diff explanation to observe its impact on the similarity of the generated CMs to those generated by OMG. In addition, we used the \textit{Diff Narrative} as a diff augmentation to determine if the additional details provided by FIDEX help the SLM write CMs more similar to those produced by OMG.
%This led the selected SLM to achieve a BLEU score of 13.01, which is 21\% higher than when only the raw diff is used. Given the high automated scores achieved by this diff augmentation, 
Lastly, we conducted a final practitioner survey to measure the effectiveness of CMG using the selected SLM. 
Figure \ref{fig:us-vs-omg} highlights the differences between OMG and our final CMG approach, OMEGA.

% \subsection{Survey}

\begin{table*}[h!]
\caption{Automated Evaluation Results for Generating \textit{LLM-derived commit context}. 
% All the models are quantized to 4 bits using the AWQ technique. 
% The best and second-best values for each RQ are bold and underlined, respectively. 
SMA stands for Software Maintenance Activity\textsuperscript{*}. We used the value reported by Li et al. to avoid incurring additional costs\textsuperscript{$\Delta$}. Bold models are the selected candidates.}
\label{table:context-validation}
\centering
\resizebox{\textwidth}{!}{
\begin{tabularx}{\textwidth}{Xcccc|ccc|c}
\toprule
\textbf{Model} & \textbf{Candidate For} &
\multicolumn{3}{c}{\textbf{Class Summarization}} & 
\multicolumn{3}{c}{\textbf{Method Summarization}} &
\textbf{SMA}\textsuperscript{*} \textbf{Classification}\\
 & & \textbf{BLEU}  & \textbf{METEOR} & \textbf{ROUGE-L} & \textbf{BLEU}  & \textbf{METEOR} & \textbf{ROUGE-L} & \textbf{Accuracy} \\
\midrule
gpt-4-turbo-2024-04-09 &  & 1.74 & 18.85 & 15.58 & 3.87 & 30.11  & 22.21 & 51\%\textsuperscript{$\Delta$} \\
\midrule
CodeFuse-DeepSeek-33B & RQ1-2 & {\ul 2.09} & 17.07 & {\ul 16.81} & {\ul 15.50} &\textbf{38.11} &\textbf{37.36} & 35\% \\
Deepseek-Coder 33B Instruct & RQ1-2 & 2.01 & 18.05 & 18.26 & 8.42 & 36.04 & 28.90 & 36\% \\
OpenCodeInterpreter-DS-33B & RQ1-2 & 1.06 & {\ul 20.21} & 16.47 &  \textbf{16.11} & {\ul 37.42} & {\ul 37.01} & {\ul 36\%}\\
\textbf{Llama3 70B Instruct} & RQ1-2 & \textbf{2.51} & \textbf{20.58} & \textbf{18.26} & 9.64 & 35.75 & 30.66 & \textbf{50\%} \\
\midrule
CodeQwen 1.5 7B Chat & RQ3 & {\ul 2.16} & \textbf{20.05} & \textbf{18.57 }& \textbf{19.46} & \textbf{38.91} & \textbf{39.58} & {\ul 37\%}\\
Mistral 7B Instruct v0.3 & RQ3 & 1.58 & {\ul 17.59} & 15.17 & 9.90 & 34.70 & 31.59 & 34\% \\
\textbf{Llama3 8B Instruct} & RQ3 & \textbf{2.18} & 16.90 & {\ul 16.76} & {\ul 13.41} & {\ul 35.66} & {\ul 33.37} & \textbf{46\%} \\
\bottomrule
\end{tabularx}
}
\end{table*}

\section{Results} \label{results}
\useunder{\uline}{\ul}{}

In this section, we address our research questions by presenting the results of both automated and human evaluations.

% \subsection{RQ1}
% \begin{boxA}
\subsection*{\textbf{RQ1.} Can an OLLM generate CMs comparable to a state-of-the-art LLM (GPT-4)?}
% \end{boxA}

% \hl{\textbf{Fixed.} In order to ensure the quality of the LLM-derived commit context, similar to Li et al.} \cite{li2024omg}, \hl{we evaluated the candidate OLLMs performance on class summary generation, MMS, and classifying software maintenance activity types. Below, we report the results.}

\begin{table*}[h!t]
\caption{Automated Evaluation Results for RQ1 and RQ2 using instruction-tuned AWQ-quantized Llama3 70B. Before and after values for commit enhancements are separated by slash. Rows with bold context were used in the survey conducted for the RQ.}
\label{table:rq1-2}
\centering
\resizebox{\textwidth}{!}{
\begin{tabularx}{\textwidth}{ccXccc|ccc}
\toprule
\textbf{RQ} & \textbf{Prompt}  & \textbf{Commit Context} &
\multicolumn{3}{c}{\textbf{Reference OMG}} & 
\multicolumn{3}{c}{\textbf{Reference Human}} \\
 & & & \textbf{BLEU}  & \textbf{METEOR} & \textbf{ROUGE-L} & \textbf{BLEU}  & \textbf{METEOR} & \textbf{ROUGE-L} \\
\midrule
\multirow{2}{*}{1} & ReAct & Same as OMG & 5.2 & 20.33 & 27.22 & 2.29 & 17.47 & 11.92 \\
 & Zero-shot & \textbf{Same as OMG} & 11.36 & 30.37 & 31.46 & 1.26 & 17.39 & 10.03 \\
\midrule
\multirow{3}{*}{2} & \multirow{3}{*}{Zero-shot} & OMG - Documentation & 10.54 / 14.12  & 29.84 / 35.78 & 27.64 / 32.76 & 1.05 / 1.09 & 14.98 / 17.04 & 7.59 / 8.76 \\
 & & MMS replaced by CMMS & 10.32 / 14.08 & 29.95 / 36.44 & 27.54 / 31.83 & 1.06 / 0.89 & 15.57 / 17.05 & 7.76 / 8.41 \\
 &  & \textbf{Refined}  & 10.54 / 14.19 & 29.84 / 36.44 & 27.64 / 32.06 & 1.05 / 0.95 & 14.98 / 16.38 & 7.59 / 8.16 \\
\bottomrule
\end{tabularx}
}
\end{table*}

\begin{table*}[h!b]
\caption{Automated Evaluation Results for RQ3 using instruction-tuned AWQ-quantized Llama3 8B. The last row presents the automated scores of OMEGA, which was used for survey 3.}
\label{table:rq3}
\centering
\resizebox{\textwidth}{!}{
\begin{tabularx}{\textwidth}{cccccc|ccc}
\toprule
\textbf{Prompt} &
\textbf{Commit Context} &
\textbf{Diff Augmentation Technique} & 
\multicolumn{3}{c}{\textbf{Reference OMG}} & 
\multicolumn{3}{c}{\textbf{Reference Human}} \\
& & & \textbf{BLEU}  & \textbf{METEOR} & \textbf{ROUGE-L} & \textbf{BLEU}  & \textbf{METEOR} & \textbf{ROUGE-L} \\
\midrule
\multirow{3}{*}{Zero-shot} & \multirow{3}{*}{Refined} & None & 10.78 & 31.18 & 28.16 & 0.86 & 14.72 & 7.38 \\
& & Diff Narrator & {\ul 12.19} & {\ul 32.26} & {\ul 29.81} &  {\ul 0.89} & {\ul 15.09} & {\ul 7.99} \\
& & \textbf{FIDEX (Used in OMEGA)} &\textbf{ 13.01} & \textbf{33.13} & \textbf{30.85} & \textbf{1.22} & \textbf{16.08} & \textbf{8.26} \\
\bottomrule
\end{tabularx}
}
\end{table*}

\subsubsection*{\textbf{Automated Evaluation}}
To ensure the quality of the LLM-derived commit context, similar to Li et al. \cite{li2024omg}, we evaluated the candidate OLLMs performance on class summarization, MMS, and classifying software maintenance activity types. 
Table \ref{table:context-validation} presents the results of automated evaluation of candidate OLLMs in generating LLM-derived commit context. 
Among the candidate OLLMs, the quantized instruction-tuned Llama3 70B scored highest in classifying software maintenance activity type and class summarization, leading to its selection for experiments in RQ1 and RQ2.
% Among the candidate OLLMs, the instruction-tuned quantized Llama3 70B scored highest in 4 out of 7 metrics, leading to its selection for experiments in RQ1 and RQ2.

As discussed in the \hyperref[rq1-prompting]{previous section}, we experimented with ReAct and zero-shot prompting techniques to answer RQ1. Table \ref{table:rq1-2} presents the automated evaluation results for these two prompting approaches.
% different prompting methods we experimented with in RQ1 
% \hl{\textbf{Done.} Explained in Secrtion X}. 
% \hl{\textbf{Removed.} Using ReAct as the prompting approach with the selected OLLM resulted in a BLEU score that was 2.18 times lower than the one achieved by the All-in-context zero-shot prompting approach. Remove this and use the following sentence after fixing. Also update methodology.} 
Based on the automated scores, using zero-shot prompting resulted in a BLEU score that was 2.18 times higher than that achieved by adopting ReAct. Therefore, we used the CMs generated through zero-shot prompting technique in our first survey to assess their quality by practitioners.

% \hl{\textbf{Added above.} All-in-context zero-shot prompting had a METEOR and ROUGE-L  score twice of the score achieved through using ReAct. So we sent these for our first practitioner survey.}

%The CMs used in our first practitioner survey achieved METEOR and ROUGE-L scores greater than 30 when compared with OMG as the reference. \hl{Add reference for 30 being good}

\subsubsection*{\textbf{Human Evaluation}}

In our first survey, 10 practitioners participated. The CMs were randomly sampled using all-in-context zero-shot prompting with the original OMG commit context. %Figure \ref{fig:survey-1} illustrates the results of this survey. 
OLLM-generated CMs were preferred in 39\% of responses, while OMG-generated CMs were selected as the overall preferred CM in 29\% of responses. 
In 52\% of the responses, the CM generated by the selected OLLM was perceived as more concise. Overall, there was no aspect in which OMG was selected by the majority of responses over the selected OLLM (quantized instruction-tuned Llama3 70B). %(OMG $<$ Identical + Selected OLLM in all aspects). 
However, OLLM-generated CMs were selected in fewer responses in terms of comprehensiveness (OLLM:25\% of responses and OMG:33\%). This observation led to the formulation of RQ2, which we addressed through LLM-derived commit context enhancements.

% \begin{figure}[]
%   \centering
%   \includegraphics[width=\linewidth]{survey-1-figure.eps}
%   \caption{Human Evaluation Results of Llama3 70B AWQ VS OMG}
%   \label{fig:survey-1}
% \end{figure}

\begin{boxH}
An OLLM can produce CMs that are comparable overall to those generated by a state-of-the-art LLM except in terms of comprehensiveness.
%State-of-the-art CMG does not need a proprietary LLM.
% However, the CMs generated using the original commit context lack the expected comprehensiveness and expressiveness.
\end{boxH}

% To select an OLLM for answering RQ1 and an SLM for answering RQ3, we measured the performance of candidate models on generating the LLM-derived commit context. Table \ref{table:context-validation} presents the results of this evaluation. Among the candidate OLLMs in RQ1, the instruction-tuned quantized Llama3 70B scored highest in 4 out of 7 metrics, leading to its selection for experiments in RQ1 and RQ2. Among the candidate SLMs, the conversation-tuned quantized CodeQwen 7B achieved the highest score in 5 out of 7 metrics. 
% \hl{I am not sure I understand what Table 1 is. trying to depict?}
% Initially, we chose this model to continue the experiments in RQ3. However, as explained in the previous section, we ultimately selected the instruction-tuned quantized Llama3 8B as the chosen SLM due to technical issues with the initial model.

\subsection*{\textbf{RQ2.} How can we bridge the comprehensiveness gap between the CMs produced by an OLLM and the CMs generated by a state-of-the-art LLM (GPT-4)?}

\subsubsection*{\textbf{Automated Evaluation}}
Table \ref{table:rq1-2} presents the automated evaluation results for our ablation study (See \hyperref[method]{Section III}), highlighting the impact of each contextual refinement in bridging the identified comprehensiveness gap from our first survey.
Removing documentation from affected method and class bodies %(346 commits)
significantly improved all automated scores, increasing the BLEU score of the CMs from 10.54 to 14.12, a 34\% improvement. Replacing MMS with CMMS %(280 commits) 
without documentation removal also led to similar improvement, although the BLEU score when comparing CMs to human-written ones degraded by 16\%. Given the effectiveness of each standalone enhancement, we combined them, resulting in a BLEU score of 14.19 when compared to OMG, a 35\% higher score than the score achieved by CMs used for our first practitioner survey.

% Initially, we chose this model to continue the experiments in RQ3. 
% However, as explained in the previous section, we ultimately selected the instruction-tuned quantized Llama3 8B as the chosen SLM due to technical issues with the initial model.

\subsubsection*{\textbf{Human Evaluation}}

Our second survey was designed to assess the effectiveness of our strategy to bridge the identified comprehensiveness gap in the first survey. Hence, 10 participants compared 15 randomly sampled CMs that were generated before the LLM-derived context enhancements with those generated by the enhanced context. Based on the survey results, 71\% of responses found the CM generated through the enhanced context more comprehensive.

\begin{boxH}
Replacing MMS with CMMS, along with a documentation removal step before summarizing affected methods and classes, mitigates the comprehensiveness gap.
\end{boxH}

\subsection*{\textbf{RQ3.} Can a smaller OLLM (SLM) produce CMs comparable to a state-of-the-art LLM (GPT-4)?}

\subsubsection*{\textbf{Automated Evaluation}}
% \hl{add intro sentence}
Similar to RQ1, we evaluated the candidate SLMs performance on class summarization, MMS, and classifying software maintenance activity types. 
Table \ref{table:context-validation} presents the automated scores achieved by each candidate SLM in generating LLM-derived commit context. 
Although the quantized conversation-tuned CodeQwen 7B was the best performing model based on the automated scores, we chose the second best SLM, the quantized instruction-tuned Llama3 8B, after observing the problem of repeated sentences by the quantized conversation-tuned CodeQwen 7B in CMG (See \hyperref[slm-selection]{previous section} for details).

% \hl{Among the candidate SLMs, we chose quantized instruction-tuned Llama3 8B
% for the answering RQ3 since conversation-tuned quantized CodeQwen 7B, in spite of being the best performer in terms of the automated metrics, often repeated the same sentences, producing lower quality CMs (See Section X for details).}
We present the automated scores for our experiments in RQ3 in Table \ref{table:rq3}. As shown in the table, each augmentation to the raw diff consistently improves all automated scores. Two key observations can be made from this table. Firstly, the FIDEX-generated diff explanation enhances the BLEU score when compared to human-written CMs by 42\%. %This improvement allowed the selected SLM to outperform the selected OLLM in terms of similarity to human-written CMs. 
Secondly, the difference between various diff augmentation approaches compared to the raw diff is more significant than the differences among these methods themselves. We posit that this is because providing the Diff Narrative gives sufficient context for simple diffs, and the CMs for those diffs do not significantly change with the FIDEX approach. Nevertheless, appending the diff with our FIDEX-generated diff explanation improves the BLEU score by 21\%, which is only 8\% lower than the score achieved by the selected OLLM without diff augmentation (RQ2).

% \begin{figure}[b]
%   \centering
%   \includegraphics[width=\linewidth]{survey-1-figure.eps}
%   \caption{Human Evaluation Results of Llama3 70B AWQ VS OMG}
%   \label{fig:survey-1}
% \end{figure}

\subsubsection*{\textbf{Human Evaluation}}

Our last survey aimed to evaluate the performance of OMEGA by assessing the impact of augmenting commit diff with FIDEX-generated explanations compared to OMG.
% Our last survey aimed to assess the impact of augmenting commit diff with FIDEX-generated diff explanation to evaluate SLMs performance compared to OMG. 
A total of 22 practitioners participated in this survey. According to the findings, the selected SLM outperforms GPT-4 in generating CMs that meet practitioners' expectations, except in conciseness. SLM-generated CMs were preferred in 46\% of responses, while OMG-generated CMs were preferred in 34\% of responses. Notably, the preference for SLM-generated CMs in all metrics, except conciseness, was higher than that for the quantized instruction-tuned Llama3 (OLLM). This highlights the effectiveness of the LLM-derived commit context enhancements (RQ2) and the FIDEX-generated diff summaries (RQ3) in enabling a quantized OLLM to produce superior CMs compared to GPT-4.

\begin{boxH}
Augmenting the commit diff with an explanation generated by FIDEX enables an SLM with just 0.005\% of the trained parameters to outperform GPT-4 in CMG.
\end{boxH}

\begin{figure}[t]
  \centering
  \includegraphics[width=\linewidth]{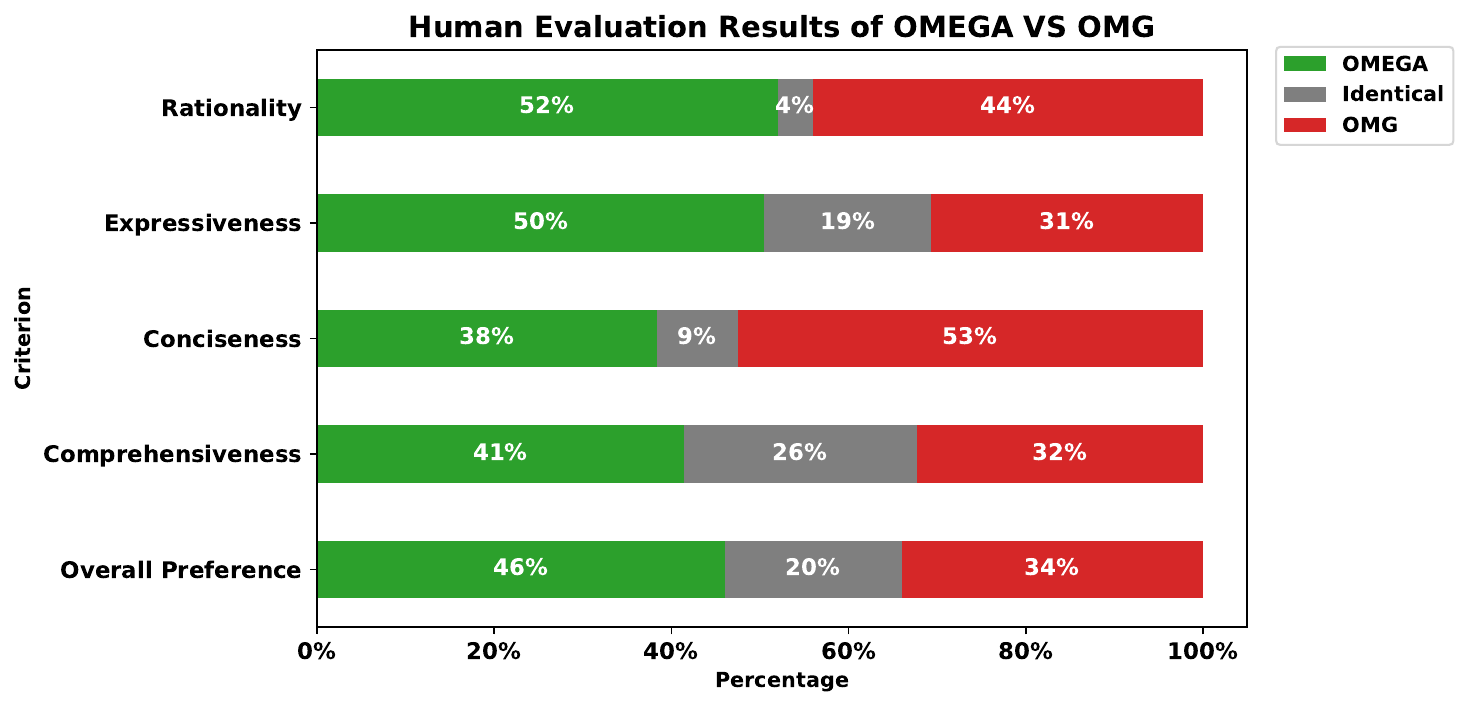}
  \caption{Human Evaluation Results of OMEGA VS OMG}
  \label{fig:survey-3}
\end{figure}

% \section{Implications} \label{implications}

% \subfile{Sections/implications}

\section{Threats to Validity} \label{threats}

We adopted measures to mitigate potential threats to the validity of our work, which we outline in this section.

\textbf{Construct Validity} 
Relying on automated machine translation metrics for CM evaluation has been shown not to align with human preferences\cite{li2024omg}. To minimize our reliance on these metrics, we used them solely as a preliminary step to ensure the generated CMs’ similarity to OMG-written ones. The quality of the CMs were ultimately assessed by practitioners using state-of-the-art human evaluation metrics \cite{li2024omg}.
There is a possibility that survey participants may have misunderstood the human evaluation criteria when assessing the candidate CMs. To mitigate this risk, we provided definitions for each human evaluation criterion at the beginning of the survey. %Additionally, each question included a button that allowed participants to view the definitions at any time during the survey. This ensured that participants could easily review the criteria if they had any doubts, thereby reducing the likelihood of misunderstandings and increasing the reliability of their assessments.
Lastly, our choice of quantization method may have hampered the selected model's performance. However, among the existing quantization methods at the time of conducting our study, our selected approach was the one with the minimal impact on the model's performance \cite{Lin2023AWQ:Acceleration}.

% \hl{\textbf{Does the new sentence help?} Not sure I understand why this one is a threat}
\textbf{Internal Validity} 
To ensure that our results solely stem from the contextual refinements we made and are not influenced by the selection of LLM inference parameters, 
% To ensure that our commit context refinements were the sole source of variation in the generated CMs, 
we set the temperature to 0 and avoided any repetition or frequency penalties. This approach makes the OLLMs' responses reproducible and deterministic. %by avoiding any creativity and preventing the penalization of tokens in the response.

\textbf{External Validity} Similar to OMG, our study is limited to Apache projects written in the Java programming language. However, given the multilingual programming datasets used to train our selected OLLMs, we posit that our results are generalizable to other programming languages as well.
Additionally, the choice of OLLM/SLMs may affect the generalizability of our findings. However, to mitigate this threat, we tested our approach with multiple candidate OLLM/SLMs before selecting the best performing one.%as well as the selected OLLM and observed consistent improvements in their automated scores. %This suggests that our context refinements are applicable to any OLLM carefully selected based on relevant LLM benchmark scores. 
%Results of these experiments are in our supplementary materials \cite{icse25replication}.

\section{Conclusion} \label{conclusion}

This study investigated the feasibility of generating state-of-the-art CMs that meet practitioners' expectations using OLLMs. Our results show that an OLLM can produce CMs comparable to those generated by proprietary models, though with poorer comprehensiveness performance. By refining the LLM-derived commit context, we bridged this performance gap. Additionally, our findings demonstrate that even a smaller LLM (SLM) can perform as well as, if not better, compared to a large proprietary LLM or large OLLM.

Our findings encourage researchers to explore ways to carefully curate task-specific contextual information for LLMs to achieve better results rather than relying solely on larger proprietary LLMs with poorly devised contexts. Our approach to generating high-quality CMs using an SLM offers significant advantages for the industry. The lower hardware requirements of SLMs make them adaptable for companies and practitioners with limited computational resources and facilitate the avoidance of sharing sensitive information with external providers.
We provide the source code and datasets that were used in our experiments in the supplementary \cite{icse25replication}.

\printbibliography

@misc{codeqwen1.5,
    title = {Code with CodeQwen1.5},
    url = {https://qwenlm.github.io/blog/codeqwen1.5/},
    author = {Qwen Team},
    month = {April},
    year = {2024}
}

@misc{zhong2024chatgptreplacestackoverflowstudy,
      title={Can ChatGPT replace StackOverflow? A Study on Robustness and Reliability of Large Language Model Code Generation}, 
      author={Li Zhong and Zilong Wang},
      year={2024},
      eprint={2308.10335},
      archivePrefix={arXiv},
}

@misc{munley2024llm4vvdevelopingllmdriventestsuite,
      title={LLM4VV: Developing LLM-Driven Testsuite for Compiler Validation}, 
      author={Christian Munley and Aaron Jarmusch and Sunita Chandrasekaran},
      year={2024},
      eprint={2310.04963},
      archivePrefix={arXiv},
}

@misc{zheng2024opencodeinterpreterintegratingcodegeneration,
      title={OpenCodeInterpreter: Integrating Code Generation with Execution and Refinement}, 
      author={Tianyu Zheng and Ge Zhang and Tianhao Shen and Xueling Liu and Bill Yuchen Lin and Jie Fu and Wenhu Chen and Xiang Yue},
      year={2024},
      eprint={2402.14658},
      archivePrefix={arXiv},
}

@misc{bigcode-evaluation-harness,
  author       = {Ben Allal, Loubna and
                  Muennighoff, Niklas and
                  Kumar Umapathi, Logesh and
                  Lipkin, Ben and
                  von Werra, Leandro},
  title = {A framework for the evaluation of code generation models},
  publisher = {GitHub},
  journal = {GitHub repository},
  howpublished = {\url{https://github.com/bigcode-project/bigcode-evaluation-harness}},
  year = 2022,
}

@misc{hou2024largelanguagemodelssoftware,
      title={Large Language Models for Software Engineering: A Systematic Literature Review}, 
      author={Xinyi Hou and Yanjie Zhao and Yue Liu and Zhou Yang and Kailong Wang and Li Li and Xiapu Luo and David Lo and John Grundy and Haoyu Wang},
      year={2024},
      eprint={2308.10620},
      archivePrefix={arXiv},
}

@misc{kong2024betterzeroshotreasoningroleplay,
      title={Better Zero-Shot Reasoning with Role-Play Prompting}, 
      author={Aobo Kong and Shiwan Zhao and Hao Chen and Qicheng Li and Yong Qin and Ruiqi Sun and Xin Zhou and Enzhi Wang and Xiaohang Dong},
      year={2024},
      eprint={2308.07702},
      archivePrefix={arXiv}
}

@inproceedings{geng2024multi-intent,
author = {Geng, Mingyang and Wang, Shangwen and Dong, Dezun and Wang, Haotian and Li, Ge and Jin, Zhi and Mao, Xiaoguang and Liao, Xiangke},
title = {Large Language Models are Few-Shot Summarizers: Multi-Intent Comment Generation via In-Context Learning},
year = {2024},
booktitle = {Proceedings of the IEEE/ACM 46th International Conference on Software Engineering},
articleno = {39},
numpages = {13}
}

@misc{fan2024exploringcapabilitiesllmscode,
      title={Exploring the Capabilities of LLMs for Code Change Related Tasks}, 
      author={Lishui Fan and Jiakun Liu and Zhongxin Liu and David Lo and Xin Xia and Shanping Li},
      year={2024},
      eprint={2407.02824},
      archivePrefix={arXiv},
}

@inproceedings{liu2024loganalysis,
author = {Liu, Yilun and Tao, Shimin and Meng, Weibin and Wang, Jingyu and Ma, Wenbing and Chen, Yuhang and Zhao, Yanqing and Yang, Hao and Jiang, Yanfei},
title = {Interpretable Online Log Analysis Using Large Language Models with Prompt Strategies},
year = {2024},
booktitle = {Proceedings of the 32nd IEEE/ACM International Conference on Program Comprehension},
pages = {35–46},
numpages = {12}
}

@inproceedings{pan2024codetranslation,
author = {Pan, Rangeet and Ibrahimzada, Ali Reza and Krishna, Rahul and Sankar, Divya and Wassi, Lambert Pouguem and Merler, Michele and Sobolev, Boris and Pavuluri, Raju and Sinha, Saurabh and Jabbarvand, Reyhaneh},
title = {Lost in Translation: A Study of Bugs Introduced by Large Language Models while Translating Code},
year = {2024},
booktitle = {Proceedings of the IEEE/ACM 46th International Conference on Software Engineering},
articleno = {82},
numpages = {13}
}

@inproceedings{vllm2023,
  title={Efficient Memory Management for Large Language Model Serving with PagedAttention},
  author={Woosuk Kwon and Zhuohan Li and Siyuan Zhuang and Ying Sheng and Lianmin Zheng and Cody Hao Yu and Joseph E. Gonzalez and Hao Zhang and Ion Stoica},
  booktitle={Proceedings of the ACM SIGOPS 29th Symposium on Operating Systems Principles},
  year={2023}
}

@misc{yin2024multitaskbasedevaluationopensourcellm,
      title={Multitask-based Evaluation of Open-Source LLM on Software Vulnerability}, 
      author={Xin Yin and Chao Ni and Shaohua Wang},
      year={2024},
      eprint={2404.02056},
      archivePrefix={arXiv},
}

@misc{deepseekcoder2024,
      title={DeepSeek-Coder: When the Large Language Model Meets Programming -- The Rise of Code Intelligence}, 
      author={Daya Guo and Qihao Zhu and Dejian Yang and Zhenda Xie and Kai Dong and Wentao Zhang and Guanting Chen and Xiao Bi and Y. Wu and Y. K. Li and Fuli Luo and Yingfei Xiong and Wenfeng Liang},
      year={2024},
      eprint={2401.14196},
      archivePrefix={arXiv},
}

@misc{xing2024understandingweaknesslargelanguage,
      title={Understanding the Weakness of Large Language Model Agents within a Complex Android Environment}, 
      author={Mingzhe Xing and Rongkai Zhang and Hui Xue and Qi Chen and Fan Yang and Zhen Xiao},
      year={2024},
      eprint={2402.06596},
      archivePrefix={arXiv}
}

@misc{liu2024repoqaevaluatinglongcontext,
      title={RepoQA: Evaluating Long Context Code Understanding}, 
      author={Jiawei Liu and Jia Le Tian and Vijay Daita and Yuxiang Wei and Yifeng Ding and Yuhan Katherine Wang and Jun Yang and Lingming Zhang},
      year={2024},
      eprint={2406.06025},
      archivePrefix={arXiv},
}

@article{li2024omg,
author = {Li, Jiawei and Farag\'{o}, David and Petrov, Christian and Ahmed, Iftekhar},
title = {Only diff Is Not Enough: Generating Commit Messages Leveraging Reasoning and Action of Large Language Model},
year = {2024},
issue_date = {July 2024},
volume = {1},
number = {FSE},
journal = {Proc. ACM Softw. Eng.},
month = {jul},
articleno = {34},
numpages = {22},
}

@INPROCEEDINGS{Eliseeva2023,
  author={Eliseeva, Aleksandra and Sokolov, Yaroslav and Bogomolov, Egor and Golubev, Yaroslav and Dig, Danny and Bryksin, Timofey},
  booktitle={2023 38th IEEE/ACM International Conference on Automated Software Engineering (ASE)}, 
  title={From Commit Message Generation to History-Aware Commit Message Completion}, 
  year={2023},
  volume={},
  number={},
  pages={723-735},
}

@inproceedings{Ferreira2023,
author = {Ferreira, M\'{\i}vian and Gon\c{c}alves, Diego and Bigonha, Mariza and Ferreira, Kecia},
title = {Characterizing Commits in Open-Source Software},
year = {2023},
booktitle = {Proceedings of the XXI Brazilian Symposium on Software Quality},
articleno = {7},
numpages = {10}
}

@misc{liu2021pretrainpromptpredictsystematic,
      title={Pre-train, Prompt, and Predict: A Systematic Survey of Prompting Methods in Natural Language Processing}, 
      author={Pengfei Liu and Weizhe Yuan and Jinlan Fu and Zhengbao Jiang and Hiroaki Hayashi and Graham Neubig},
      year={2021},
      eprint={2107.13586},
      archivePrefix={arXiv},
      primaryClass={cs.CL},
}

@inproceedings{du2024,
author = {Du, Xueying and Liu, Mingwei and Wang, Kaixin and Wang, Hanlin and Liu, Junwei and Chen, Yixuan and Feng, Jiayi and Sha, Chaofeng and Peng, Xin and Lou, Yiling},
title = {Evaluating Large Language Models in Class-Level Code Generation},
year = {2024},
booktitle = {Proceedings of the IEEE/ACM 46th International Conference on Software Engineering},
articleno = {81},
numpages = {13}
}

@INPROCEEDINGS{dong2023,
  author={Dong, Jinhao and Lou, Yiling and Hao, Dan and Tan, Lin},
  booktitle={2023 IEEE/ACM 45th International Conference on Software Engineering (ICSE)}, 
  title={Revisiting Learning-based Commit Message Generation}, 
  year={2023},
  volume={},
  number={},
  pages={794-805},
}

@misc{openai2024chatgpt,
    author = {OpenAI},
    title = {ChatGPT},
    year = {2024},
    url = {https://www.openai.com/research/chatgpt},
    note = {Accessed: 2024-07-28}
}

@INPROCEEDINGS{song2022,
  author={Song, Zixuan and Shang, Xiuwei and Li, Mengxuan and Chen, Rong and Li, Hui and Guo, Shikai},
  booktitle={2022 IEEE 13th International Symposium on Parallel Architectures, Algorithms and Programming (PAAP)}, 
  title={Do Not Have Enough Data? An Easy Data Augmentation for Code Summarization}, 
  year={2022},
  volume={},
  number={},
  pages={1-6},
}

@misc{moghaddam2023boostingtheoryofmindperformancelarge,
      title={Boosting Theory-of-Mind Performance in Large Language Models via Prompting}, 
      author={Shima Rahimi Moghaddam and Christopher J. Honey},
      year={2023},
      eprint={2304.11490},
      archivePrefix={arXiv},
}

@inproceedings{Tian2022WhatMessage,
author = {Tian, Yingchen and Zhang, Yuxia and Stol, Klaas-Jan and Jiang, Lin and Liu, Hui},
title = {What makes a good commit message?},
year = {2022},
booktitle = {Proceedings of the 44th International Conference on Software Engineering},
pages = {2389–2401},
numpages = {13},
}

@inproceedings{Dong2022FIRA:Generation,
author = {Dong, Jinhao and Lou, Yiling and Zhu, Qihao and Sun, Zeyu and Li, Zhilin and Zhang, Wenjie and Hao, Dan},
title = {FIRA: Fine-grained graph-based code change representation for automated commit message generation},
year = {2022},
booktitle = {Proceedings of the 44th International Conference on Software Engineering},
pages = {970–981},
numpages = {12},
}

@misc{ziegler2020finetuninglanguagemodelshuman,
      title={Fine-Tuning Language Models from Human Preferences}, 
      author={Daniel M. Ziegler and Nisan Stiennon and Jeffrey Wu and Tom B. Brown and Alec Radford and Dario Amodei and Paul Christiano and Geoffrey Irving},
      year={2020},
      eprint={1909.08593},
      archivePrefix={arXiv},
}

@ARTICLE{zhang2024automatic-cmg-crititical-review,
  author={Zhang, Yuxia and Qiu, Zhiqing and Stol, Klaas-Jan and Zhu, Wenhui and Zhu, Jiaxin and Tian, Yingchen and Liu, Hui},
  journal={IEEE Transactions on Software Engineering}, 
  title={Automatic Commit Message Generation: A Critical Review and Directions for Future Work}, 
  year={2024},
  volume={50},
  number={4},
  pages={816-835},
 }

@misc{jiang2023mistral7b,
      title={Mistral 7B}, 
      author={Albert Q. Jiang and Alexandre Sablayrolles and Arthur Mensch and Chris Bamford and Devendra Singh Chaplot and Diego de las Casas and Florian Bressand and Gianna Lengyel and Guillaume Lample and Lucile Saulnier and Lélio Renard Lavaud and Marie-Anne Lachaux and Pierre Stock and Teven Le Scao and Thibaut Lavril and Thomas Wang and Timothée Lacroix and William El Sayed},
      year={2023},
      eprint={2310.06825},
      archivePrefix={arXiv},
}

@inproceedings{galindo2023,
author = {Galindo, Jos\'{e} A. and Dominguez, Antonio J. and White, Jules and Benavides, David},
title = {Large Language Models to generate meaningful feature model instances},
year = {2023},
booktitle = {Proceedings of the 27th ACM International Systems and Software Product Line Conference - Volume A},
pages = {15–26},
numpages = {12}
}

@inproceedings{wohlin2014snowball,
author = {Wohlin, Claes},
title = {Guidelines for snowballing in systematic literature studies and a replication in software engineering},
year = {2014},
booktitle = {Proceedings of the 18th International Conference on Evaluation and Assessment in Software Engineering},
articleno = {38},
numpages = {10}
}

@inproceedings{multi-intent-2024,
author = {Geng, Mingyang and Wang, Shangwen and Dong, Dezun and Wang, Haotian and Li, Ge and Jin, Zhi and Mao, Xiaoguang and Liao, Xiangke},
title = {Large Language Models are Few-Shot Summarizers: Multi-Intent Comment Generation via In-Context Learning},
year = {2024},
booktitle = {Proceedings of the IEEE/ACM 46th International Conference on Software Engineering},
articleno = {39},
numpages = {13}
}

@misc{shi2022raceretrievalaugmentedcommitmessage,
      title={RACE: Retrieval-Augmented Commit Message Generation}, 
      author={Ensheng Shi and Yanlin Wang and Wei Tao and Lun Du and Hongyu Zhang and Shi Han and Dongmei Zhang and Hongbin Sun},
      year={2022},
      eprint={2203.02700},
      archivePrefix={arXiv},
}

@misc{icse25replication,
    title = {{OMEGA Repository}},
    url = {https://github.com/aaron-imani/omega}
}

@techreport{gpt4,
    title = {{GPT-4 Technical Report}},
    year = {2023},
    author = {{OpenAI} and Achiam, Josh and Adler, Steven and Agarwal, Sandhini and Ahmad, Lama and Akkaya, Ilge and Aleman, Florencia Leoni and Almeida, Diogo and Altenschmidt, Janko and Altman, Sam and Anadkat, Shyamal and Avila, Red and Babuschkin, Igor and Balaji, Suchir and Balcom, Valerie and Baltescu, Paul and Bao, Haiming and Bavarian, Mohammad and Belgum, Jeff and Bello, Irwan and Berdine, Jake and Bernadett-Shapiro, Gabriel and Berner, Christopher and Bogdonoff, Lenny and Boiko, Oleg and Boyd, Madelaine and Brakman, Anna-Luisa and Brockman, Greg and Brooks, Tim and Brundage, Miles and Button, Kevin and Cai, Trevor and Campbell, Rosie and Cann, Andrew and Carey, Brittany and Carlson, Chelsea and Carmichael, Rory and Chan, Brooke and Chang, Che and Chantzis, Fotis and Chen, Derek and Chen, Sully and Chen, Ruby and Chen, Jason and Chen, Mark and Chess, Ben and Cho, Chester and Chu, Casey and Chung, Hyung Won and Cummings, Dave and Currier, Jeremiah and Dai, Yunxing and Decareaux, Cory and Degry, Thomas and Deutsch, Noah and Deville, Damien and Dhar, Arka and Dohan, David and Dowling, Steve and Dunning, Sheila and Ecoffet, Adrien and Eleti, Atty and Eloundou, Tyna and Farhi, David and Fedus, Liam and Felix, Niko and Fishman, Simón Posada and Forte, Juston and Fulford, Isabella and Gao, Leo and Georges, Elie and Gibson, Christian and Goel, Vik and Gogineni, Tarun and Goh, Gabriel and Gontijo-Lopes, Rapha and Gordon, Jonathan and Grafstein, Morgan and Gray, Scott and Greene, Ryan and Gross, Joshua and Gu, Shixiang Shane and Guo, Yufei and Hallacy, Chris and Han, Jesse and Harris, Jeff and He, Yuchen and Heaton, Mike and Heidecke, Johannes and Hesse, Chris and Hickey, Alan and Hickey, Wade and Hoeschele, Peter and Houghton, Brandon and Hsu, Kenny and Hu, Shengli and Hu, Xin and Huizinga, Joost and Jain, Shantanu and Jain, Shawn and Jang, Joanne and Jiang, Angela and Jiang, Roger and Jin, Haozhun and Jin, Denny and Jomoto, Shino and Jonn, Billie and Jun, Heewoo and Kaftan, Tomer and Kaiser, Łukasz and Kamali, Ali and Kanitscheider, Ingmar and Keskar, Nitish Shirish and Khan, Tabarak and Kilpatrick, Logan and Kim, Jong Wook and Kim, Christina and Kim, Yongjik and Kirchner, Jan Hendrik and Kiros, Jamie and Knight, Matt and Kokotajlo, Daniel and Kondraciuk, Łukasz and Kondrich, Andrew and Konstantinidis, Aris and Kosic, Kyle and Krueger, Gretchen and Kuo, Vishal and Lampe, Michael and Lan, Ikai and Lee, Teddy and Leike, Jan and Leung, Jade and Levy, Daniel and Li, Chak Ming and Lim, Rachel and Lin, Molly and Lin, Stephanie and Litwin, Mateusz and Lopez, Theresa and Lowe, Ryan and Lue, Patricia and Makanju, Anna and Malfacini, Kim and Manning, Sam and Markov, Todor and Markovski, Yaniv and Martin, Bianca and Mayer, Katie and Mayne, Andrew and McGrew, Bob and McKinney, Scott Mayer and McLeavey, Christine and McMillan, Paul and McNeil, Jake and Medina, David and Mehta, Aalok and Menick, Jacob and Metz, Luke and Mishchenko, Andrey and Mishkin, Pamela and Monaco, Vinnie and Morikawa, Evan and Mossing, Daniel and Mu, Tong and Murati, Mira and Murk, Oleg and M{\'{e}}ly, David and Nair, Ashvin and Nakano, Reiichiro and Nayak, Rajeev and Neelakantan, Arvind and Ngo, Richard and Noh, Hyeonwoo and Ouyang, Long and O'Keefe, Cullen and Pachocki, Jakub and Paino, Alex and Palermo, Joe and Pantuliano, Ashley and Parascandolo, Giambattista and Parish, Joel and Parparita, Emy and Passos, Alex and Pavlov, Mikhail and Peng, Andrew and Perelman, Adam and Peres, Filipe de Avila Belbute and Petrov, Michael and Pinto, Henrique Ponde de Oliveira and {Michael} and {Pokorny} and Pokrass, Michelle and Pong, Vitchyr H. and Powell, Tolly and Power, Alethea and Power, Boris and Proehl, Elizabeth and Puri, Raul and Radford, Alec and Rae, Jack and Ramesh, Aditya and Raymond, Cameron and Real, Francis and Rimbach, Kendra and Ross, Carl and Rotsted, Bob and Roussez, Henri and Ryder, Nick and Saltarelli, Mario and Sanders, Ted and Santurkar, Shibani and Sastry, Girish and Schmidt, Heather and Schnurr, David and Schulman, John and Selsam, Daniel and Sheppard, Kyla and Sherbakov, Toki and Shieh, Jessica and Shoker, Sarah and Shyam, Pranav and Sidor, Szymon and Sigler, Eric and Simens, Maddie and Sitkin, Jordan and Slama, Katarina and Sohl, Ian and Sokolowsky, Benjamin and Song, Yang and Staudacher, Natalie and Such, Felipe Petroski and Summers, Natalie and Sutskever, Ilya and Tang, Jie and Tezak, Nikolas and Thompson, Madeleine B. and Tillet, Phil and Tootoonchian, Amin and Tseng, Elizabeth and Tuggle, Preston and Turley, Nick and Tworek, Jerry and Uribe, Juan Felipe Cerón and Vallone, Andrea and Vijayvergiya, Arun and Voss, Chelsea and Wainwright, Carroll and Wang, Justin Jay and Wang, Alvin and Wang, Ben and Ward, Jonathan and Wei, Jason and Weinmann, CJ and Welihinda, Akila and Welinder, Peter and Weng, Jiayi and Weng, Lilian and Wiethoff, Matt and Willner, Dave and Winter, Clemens and Wolrich, Samuel and Wong, Hannah and Workman, Lauren and Wu, Sherwin and Wu, Jeff and Wu, Michael and Xiao, Kai and Xu, Tao and Yoo, Sarah and Yu, Kevin and Yuan, Qiming and Zaremba, Wojciech and Zellers, Rowan and Zhang, Chong and Zhang, Marvin and Zhao, Shengjia and Zheng, Tianhao and Zhuang, Juntang and Zhuk, William and Zoph, Barret},
    month = {3},
    arxivId = {2303.08774}
}

@misc{llama3,
    title = {{Meta Llama 3}},
    url = {https://llama.meta.com/llama3/}
}

@article{Yao2024AUgly,
    title = {{A survey on large language model (LLM) security and privacy: The Good, The Bad, and The Ugly}},
    year = {2024},
    journal = {High-Confidence Computing},
    author = {Yao, Yifan and Duan, Jinhao and Xu, Kaidi and Cai, Yuanfang and Sun, Zhibo and Zhang, Yue},
    number = {2},
    pages = {100211},
    volume = {4}
}

@misc{AmazonsWarn,
    title = {{Amazon’s Q has ‘severe hallucinations’ and leaks confidential data in public preview, employees warn}},
    url = {https://www.platformer.news/amazons-q-has-severe-hallucinations/}
}

@article{Liu2022ATOM:Ranking,
    title = {{ATOM: Commit Message Generation Based on Abstract Syntax Tree and Hybrid Ranking}},
    year = {2022},
    journal = {IEEE Transactions on Software Engineering},
    author = {Liu, Shangqing and Gao, Cuiyun and Chen, Sen and Nie, Lun Yiu and Liu, Yang},
    number = {5},
    pages = {1800--1817},
    volume = {48}
}

@article{Lin2023AWQ:Acceleration,
    title = {{AWQ: Activation-aware Weight Quantization for LLM Compression and Acceleration}},
    year = {2023},
    author = {Lin, Ji and Tang, Jiaming and Tang, Haotian and Yang, Shang and Chen, Wei-Ming and Wang, Wei-Chen and Xiao, Guangxuan and Dang, Xingyu and Gan, Chuang and Han, Song},
    month = {6},
    arxivId = {2306.00978}
}

@inproceedings{He2023COME:Embedding,
    title = {{COME: Commit Message Generation with Modification Embedding}},
    year = {2023},
    booktitle = {Proceedings of the 32nd ACM SIGSOFT International Symposium on Software Testing and Analysis},
    author = {He, Yichen and Wang, Liran and Wang, Kaiyi and Zhang, Yupeng and Zhang, Hang and Li, Zhoujun},
    pages = {792--803},
    series = {ISSTA 2023},
    publisher = {},
    address = {}
}

@misc{GPT-4Wikipedia,
    title = {{GPT-4 - Wikipedia}},
    url = {https://en.wikipedia.org/wiki/GPT-4}
}

@misc{DaiGPU-Benchmarks-on-LLM-Inference,
    title = {{GPU-Benchmarks-on-LLM-Inference}},
    author = {Dai, Jack},
    url = {https://github.com/XiongjieDai/GPU-Benchmarks-on-LLM-Inference}
}

@inproceedings{Liu2023IsGeneration,
    title = {{Is Your Code Generated by ChatGPT Really Correct? Rigorous Evaluation of Large Language Models for Code Generation}},
    year = {2023},
    booktitle = {Advances in Neural Information Processing Systems},
    author = {Liu, Jiawei and Xia, Chunqiu Steven and Wang, Yuyao and ZHANG, LINGMING},
    pages = {21558--21572},
    volume = {36},
    publisher = {}
}

@article{Tao2024KADEL:Generation,
    title = {{KADEL: Knowledge-Aware Denoising Learning for Commit Message Generation}},
    year = {2024},
    journal = {ACM Trans. Softw. Eng. Methodol.},
    author = {Tao, Wei and Zhou, Yucheng and Wang, Yanlin and Zhang, Hongyu and Wang, Haofen and Zhang, Wenqiang},
    number = {5},
    month = {6},
    volume = {33},
    publisher = {Association for Computing Machinery},
    address = {New York, NY, USA}
}

@misc{LangChain,
    title = {{LangChain}},
    url = {https://www.langchain.com/}
}

@misc{LlamaParseur,
    title = {{Llama 3 performance and cost benchmarks | Parseur{\textregistered}}},
    url = {https://parseur.com/blog/blog-llama3-performance-cost}
}

@article{Faiz2023LLMCarbon:Models,
    title = {{LLMCarbon: Modeling the end-to-end Carbon Footprint of Large Language Models}},
    year = {2023},
    author = {Faiz, Ahmad and Kaneda, Sotaro and Wang, Ruhan and Osi, Rita and Sharma, Prateek and Chen, Fan and Jiang, Lei},
    month = {9},
    arxivId = {2309.14393}
}

@article{Wang2024Multi-grainedGeneration,
    title = {{Multi-grained contextual code representation learning for commit message generation}},
    year = {2024},
    journal = {Information and Software Technology},
    author = {Wang, Chuangwei and Zhang, Li and Zhang, Xiaofang},
    pages = {107393},
    volume = {167}
}

@inproceedings{Tao2021OnStudy,
    title = {{On the Evaluation of Commit Message Generation Models: An Experimental Study}},
    year = {2021},
    booktitle = {2021 IEEE International Conference on Software Maintenance and Evolution (ICSME)},
    author = {Tao, Wei and Wang, Yanlin and Shi, Ensheng and Du, Lun and Han, Shi and Zhang, Hongyu and Zhang, Dongmei and Zhang, Wenqiang},
    pages = {126--136}
}

@misc{QuestionPro,
    title = {{QuestionPro}},
    url = {https://www.questionpro.com}
}

@article{Yao2022ReAct:Models,
    title = {{ReAct: Synergizing Reasoning and Acting in Language Models}},
    year = {2022},
    author = {Yao, Shunyu and Zhao, Jeffrey and Yu, Dian and Du, Nan and Shafran, Izhak and Narasimhan, Karthik and Cao, Yuan},
    month = {10},
    arxivId = {2210.03629}
}

@article{Rebai2020RecommendingAnalysis,
    title = {{Recommending refactorings via commit message analysis}},
    year = {2020},
    journal = {Information and Software Technology},
    author = {Rebai, Soumaya and Kessentini, Marouane and Alizadeh, Vahid and Sghaier, Oussama Ben and Kazman, Rick},
    pages = {106332},
    volume = {126}
}

@inproceedings{Chien2023Reducing2035,
    title = {{Reducing the Carbon Impact of Generative AI Inference (today and in 2035)}},
    year = {2023},
    booktitle = {Proceedings of the 2nd Workshop on Sustainable Computer Systems},
    author = {Chien, Andrew A and Lin, Liuzixuan and Nguyen, Hai and Rao, Varsha and Sharma, Tristan and Wijayawardana, Rajini},
    series = {HotCarbon '23},
    publisher = {Association for Computing Machinery},
    address = {New York, NY, USA}
}

@article{Liu2024RepoQA:Understanding,
    title = {{RepoQA: Evaluating Long Context Code Understanding}},
    year = {2024},
    author = {Liu, Jiawei and Tian, Jia Le and Daita, Vijay and Wei, Yuxiang and Ding, Yifeng and Wang, Yuhan Katherine and Yang, Jun and Zhang, Lingming},
    month = {6},
    arxivId = {2406.06025}
}

@misc{SamsungLeak,
    title = {{Samsung Bans ChatGPT Among Employees After Sensitive Code Leak}},
    url = {https://tinyurl.com/samsung-leaks}
}
% \bibliography{references}

\end{document}